\documentclass[a4paper,fleqn,usenatbib,useAMS]{mnras}

\usepackage{newtxtext,newtxmath}

\usepackage[T1]{fontenc}
\usepackage{ae,aecompl}

\usepackage{graphicx}	
\usepackage{amsmath}	
\usepackage{amssymb}	
\usepackage{lscape}
\usepackage{ulem}

\newcommand{\msun}{M_{\odot}}
\newcommand{\rhalf}{R_{1/2}}
\newcommand{\rsep}{r_{\rm sep}}
\newcommand{\rhalfh}{R^{\rm host}_{1/2}}
\newcommand{\rhalfc}{R^{\rm comp}_{1/2}}

\newcommand{\rsepdm}{r_{\rm sep}^{\rm DM}}

\newcommand{\new}[1]{#1}

\title[Interacting Galaxies in IllustrisTNG - I] 
{Interacting galaxies in the IllustrisTNG simulations - I: Triggered star formation in a cosmological context}

\author[D. R. Patton et al.]{
David R. Patton,$^{1}$\thanks{E-mail: dpatton@trentu.ca}
Kieran D. Wilson$^{1}$,
Colin J. Metrow$^{1}$,
Sara L. Ellison$^{2}$,\newauthor
Paul  Torrey$^{3}$, 
Westley Brown$^{1}$,
Maan H. Hani$^{2}$,
Stuart McAlpine$^{4}$,  
Jorge Moreno$^{5}$\newauthor  
and Joanna Woo$^{2}$  
\\
$^{1}$Department of Physics and Astronomy, Trent University, 1600 West Bank Drive, Peterborough, ON K9L 0G2, Canada\\
$^{2}$Department of Physics and Astronomy, University of Victoria, Finnerty Road, Victoria, BC V8P 1A1, Canada\\
$^{3}$Department of Astronomy, University of Florida, 211 Bryant Space Science Center, Gainesville, FL 32611, USA\\
$^{4}$Department of Physics, University of Helsinki, Gustaf H\"allstr\"omin katu 2a P.O. Box 64, FI-00014 University of Helsinki, Finland\\
$^{5}$Department of Physics and Astronomy, Pomona College, Claremont, CA 91711, USA\\
}

\pubyear{2016}

\begin{document}
\label{firstpage}
\pagerange{\pageref{firstpage}--\pageref{lastpage}}
\maketitle

\begin{abstract}

We use the IllustrisTNG cosmological hydrodynamical simulations 
to  investigate how the specific star formation rates (sSFRs) of massive galaxies ($M_* > 10^{10} \msun$)
depend on the distance to their closest companions.  
We estimate sSFR enhancements by comparing with control samples that are matched in redshift, 
stellar mass, local density and isolation, \new{and we restrict our analysis to pairs with stellar mass ratios of 0.1 to 10}.  
At small separations ($\sim$ 15 kpc), the mean sSFR is enhanced by a factor of $2.0 \pm 0.1$ in the flagship (110.7 Mpc)$^3$ simulation (TNG100-1).  
Statistically significant enhancements extend out to 3D separations of 280 kpc in the (302.6 Mpc)$^3$ simulation (TNG300-1).  
We find similar trends in the EAGLE and Illustris simulations, although their sSFR enhancements are lower than those in TNG100-1 
by about a factor of two.  
Enhancements in IllustrisTNG galaxies are seen throughout the redshift range explored ($0 \leq z < 1$), 
with the strength of the enhancements decreasing with increasing redshift for galaxies with close companions.
In order to more closely compare with observational results, we separately consider 2D projected distances between galaxies in IllustrisTNG.
We detect significant sSFR enhancements out to projected separations of 260 kpc in TNG300-1, 
with projection effects diluting the size of the enhancements by about 20 per cent below 50 kpc.
We find similar sSFR enhancements in TNG100-1 and Sloan Digital Sky Survey galaxies, 
with enhancements extending out to projected separations of about 150 kpc for star-forming galaxies at $z < 0.2$.
Finally, by summing over all separations, we estimate that the presence of closest companions 
boosts the average sSFR of massive galaxies in TNG100-1 by 14.5 per cent.

\end{abstract}




\section{Introduction}\label{secintro}

Interactions and mergers of galaxies play an important role in the formation and 
evolution of galaxies within the $\Lambda$CDM paradigm.   
Idealised binary merger simulations predict that galaxy properties undergo 
profound changes throughout the merger sequence, with the strength 
and nature of these changes depending on the orbital properties and mass ratio of the galaxy pair, 
and on the underlying properties of the galaxies themselves \citep{dimatteo07,cox08,lotz08,chilingarian10,moreno15}.
Merger simulations predict that interactions can trigger gas inflows, which lead to enhanced star formation, 
diluted gas-phase metallicities and increased AGN activity 
\citep{mihos96,dimatteo05,rupke10a,torrey12,hopkins13,blumenthal18,moreno19}.
These changes are often most apparent during and shortly after \new{coalescence}, 
but they are also seen earlier in the merger sequence, following close encounters between the galaxies.  

Observational studies of galaxies have provided general support for these predictions, particularly at low redshift.  
For example, studies of galaxy pairs have shown that galaxies with close companions have 
enhanced star formation \citep{barton00,ellison08,woods10,scudder12,patton13,cao16}, 
diluted metallicities \citep{ellison08,kewley10,rupke10b,scudder12,bustamante20}, 
increased structural asymmetries \citep{hernandez05,patton05,depropris07,casteels14,patton16} 
and higher AGN fractions \citep{alonso07,ellison11,satyapal14,weston17,ellison19} when compared with control samples of relatively isolated galaxies.

However, the link between merger simulations and observations of galaxy pairs is indirect.  
Idealised binary merger simulations assume pre-selected orbits and orientations, 
and span a limited combination of stellar masses, gas fractions, bulge fractions, etc.
In addition, most merger simulations do not include a circumgalactic medium or external gas accretion, 
nor do they include neighbouring galaxies or realistic group/cluster environments.
The advent of relatively high resolution cosmological zoom-in simulations 
has allowed for some progress in studying mergers within a more realistic cosmological context 
\citep{sparre16,bustamante18}; however, the number of merging systems studied in this manner 
is too small to cover a representative sample of interacting and merging galaxies.  

Observationally, galaxy pairs have been found in a wide range of environments 
\citep{barton07,mcintosh08,ellison10,lin10,patton16}.  However, the interpretation of these galaxy pair samples 
is limited by projection effects, wherein the 3D separations and relative velocities of individual pairs 
are unknown \citep{kitzbichler08}.
As such, the orbital histories of most galaxy pairs are unconstrained.  
Moreover, with each pair seen at a single epoch, information about 
the evolutionary and orbital histories of the galaxies is very limited.

Fortunately, the availability of increasingly realistic cosmological hydrodynamical simulations 
provides a way forward on many of these issues.  Simulations such as 
Illustris \citep{vogelsberger14a}, 
Horizon-AGN \citep{dubois14}, 
EAGLE \citep{schaye15}, 
MassiveBlack-II \citep{khandai15}, 
SIMBA \citep{dave19}
and IllustrisTNG \citep{nelson19} yield thousands of massive galaxies within volumes spanning at least (100 Mpc)$^3$, 
with properties that are well matched to galaxy populations at low redshift.  
Galaxy interactions and mergers arise naturally within these simulations, 
with a full distribution of orbits, spin-orbit orientations, environments and galaxy properties.

While these cosmological simulations are naturally of much lower mass resolution than typical merger simulations, 
the presence of tidal tails and star forming regions in synthetic images \citep{torrey15,pearson19} suggests that 
they may nevertheless capture triggered star formation in interacting galaxies.
Indeed, \citet{martin17} find that Horizon-AGN galaxies that are within 1 Gyr of a merger have SFRs that are enhanced by a factor of $\sim$ 1.7, 
while \citet{rodriguezmontero19} find that recently merged galaxies in the SIMBA simulations have sSFRs 
that are enhanced by a factor of 2-3.

In this study, we aim to bridge the gap between merger simulations and observations of galaxy pairs 
by identifying galaxies and their closest companions within the IllustrisTNG cosmological hydrodynamical simulations, 
focussing on the sSFRs of galaxies and the degree to which they are enhanced 
due to interactions.   By identifying galaxy pairs from these simulations using both 3D and projected separation, 
our goal is to facilitate a direct comparison with observed galaxy pairs.  \new{In a companion paper \citep{hani20}, 
we extend this analysis to the sSFRs of post-merger galaxies in IllustrisTNG.}

In the following section, we describe the simulations and the methodology used to identify the closest companion of each massive 
galaxy in the simulations.  In Section 3, we show how the sSFR and its enhancement depends on the 3D distance to 
the closest companion in the highest resolution and largest volume IllustrisTNG simulations.  In Section 4, we investigate 
how these results depend on resolution and volume within IllustrisTNG, and we carry out a direct comparison with the 
Illustris-1 and EAGLE simulations.  We then show in Section 5 how IllustrisTNG sSFR enhancements depend on projected separation, 
and we make a direct comparison with sSFR enhancements in the Sloan Digital Sky Survey (hereafter SDSS).  
We discuss our findings in Section 6 and summarise our conclusions in Section 7.

\section{Simulations and methodology}\label{secmethods}


\subsection{The IllustrisTNG simulations}

Our study is primarily focused on the cosmological hydrodynamical simulations 
from the IllustrisTNG\footnote{http://www.tng-project.org} suite.  
IllustrisTNG is the successor to the Illustris simulations (see below), with numerous updates 
that are described in the public data release \citep{nelson19}.  
\citet{weinberger18} and \citet{pillepich18a} provide a complete description of the physical models.
Additional details about these simulations are given by \citet{pillepich18b}, 
\citet{springel18}, \citet{nelson18}, \citet{naiman18}, and \citet{marinacci18}.

The IllustrisTNG simulations are publicly available for cubic periodic volumes of co-moving side length 110.7 Mpc and 302.6 Mpc.  
For each volume, baryonic and dark-matter only simulations are available at three mass resolution levels.  
The baryonic simulations follow the evolution of dark matter, stars, gas and supermassive black holes from 
a redshift of 127 to the present epoch using the AREPO moving-mesh code \citep{springel10}. 
Star formation occurs above a threshold density criterion, and is regulated by a subgrid model for the 
interstellar medium \citep{springel03,nelson15,pillepich18a}.  
Where needed for stellar evolution and mass return calculations, 
a \citet{chabrier03} initial mass function is used.

Our primary focus is on the highest resolution run for the (110.7 Mpc)$^3$ volume (TNG100-1) 
and on the highest resolution run of the (302.6 Mpc)$^3$ volume (TNG300-1).  
However, we also use two of the lower resolution simulations (TNG100-2 and TNG100-3) to assess how our primary results 
depend on resolution.  Subhalo catalogues are available at 100 redshifts, including 50 snapshots at $z < 1$.   
\new{Following \citet{nelson19}, we exclude {\textsc SUBFIND} subhalos for which the SubhaloFlag parameter has been set to 0,
as they are known to be sub-galactic clumps and are not considered to be bona fide galaxies.}
Properties of these simulations (and others described below) are summarised in Table~\ref{tabsim}.

\subsection{The Illustris-1 simulation}

While this paper focusses on the IllustrisTNG simulations, we also carry out a direct comparison with the 
original Illustris simulations \citep{vogelsberger14a}.  
\citet{nelson15} describe the Illustris public data release, 
while \citet{genel14} and \citet{vogelsberger14b} provide additional information about these simulations.
As with IllustrisTNG, the Illustris simulations use the AREPO moving-mesh code \citep{springel10}.
The physical models are described by \citet{vogelsberger13}.
For this study, we consider only the flagship Illustris-1 simulation, which is the 
highest resolution (106.5 Mpc)$^3$ hydrodynamical simulation from Illustris.  As seen in Table~\ref{tabsim}, 
Illustris-1 is comparable to TNG100-1 in terms of both volume and resolution. 
Illustris-1 has 51 snapshots at $z < 1$, with time sampling that is very similar to IllustrisTNG.

\subsection{The EAGLE simulation}

Given the inherent connections between Illustris and IllustrisTNG, we carry out an additional comparison with the
largest simulation from the {\it Evolution and Assembly of GaLaxies and their Environments} (EAGLE) project.
This simulation, denoted Ref-L0100N1504, is comparable to both TNG100-1 and Illustris-1 in terms of volume 
and mass resolution (see Table~\ref{tabsim}).
However, EAGLE is a smooth particle hydrodynamics code and does not use an adaptive mesh, and, more notably, 
adopts a considerably different subgrid scheme.  
Galaxy catalogues and particle properties are publicly available for 29 snapshots at $z < 20$.  
We instead use galaxy catalogues from a set of 201 ``snipshots'', including 65 snipshots at $z < 1$.  
\new{Following \citet{mcalpine16}, we exclude subhalos for which the Spurious flag has been set to 1,
as these objects are known to be dense concentrations within subhalos and therefore are artefacts of the
{\textsc SUBFIND} algorithm.}
For a full description of the simulation and calibration strategy, see \citet{schaye15} and \citet{crain15}, respectively.  

We briefly outline the model governing star formation here, as it is the most important model related to this study.  
Star formation occurs above the metallicity-dependent threshold of \citet{schaye04}, 
whereby `cold' gas particles ($T\sim 10^4$ K) are stochastically converted into star particles 
at a pressure-dependent rate that reproduces the observed
Kennicutt-Schmidt star formation law \citep[see][]{schaye08}.
Each newly formed star particle represents a simple stellar population with a \citet{chabrier03} initial mass function. 
The feedback from massive stars and supernovae is injected thermally and stochastically by raising the surrounding gas 
to a fixed temperature increment as described in \citet{dallavecchia12}.

\begin{table}
\centering
\caption{A summary of the simulations used in this analysis, 
including the co-moving side length of the simulation box ($L_{\rm box}$), 
the baryonic particle mass ($M_{\rm baryon}$), the average number of massive host galaxies ($M_* > 10^{10}\msun$) 
per snapshot ($\langle N^{\rm host}\rangle$), 
the average time between snapshots ($\Delta t_{\rm snap}$) at $z \lesssim 1$, 
and the 3D separation at which the crowding fraction reaches 20 per cent ($r_{\rm crowd}$).
\label{tabsim}}
\begin{tabular}{cccccc}
\hline
Simulation&$L_{\rm box}$&$M_{\rm baryon}$&$\langle N^{\rm host}\rangle$&$\Delta t_{\rm snap}$&$r_{\rm crowd}$\\
Name&(Mpc)&($\msun$)&&(Myr)&(kpc)\\
\hline
TNG100-1&110.7&$1.4\times 10^6$&5976&162&15.9\\
TNG100-2&110.7&$1.1\times 10^7$&4382&162&17.7\\
TNG100-3&110.7&$9.0\times 10^7$&2534&162&23.4\\
TNG300-1&302.6&$1.1\times 10^7$&90279&162&17.4\\
Illustris-1&106.5&$1.6\times 10^6$&5789&155&18.0\\
EAGLE&100.0&$1.8\times 10^6$&3521&121&16.3\\ 
\hline
\end{tabular}
\end{table}

\subsection{Complete samples of galaxies with $M_* > 10^9 \msun$}\label{secmass}

For each of the simulations under consideration, we begin by constructing a sample of galaxies 
which is complete above a stellar mass of $10^9 \msun$.  
Galaxies in this regime are resolved  with more than 500 stellar particles per galaxy for the highest 
resolution 100 Mpc volumes (TNG100-1, Illustris-1 and EAGLE) and at least 90 stellar particles per galaxy 
for the lower resolution TNG300-1 simulation (see Table~\ref{tabsim}).

However, in cases where two galaxies are close to one another, it can be difficult to correctly disentangle their stellar 
mass.  For all of the simulations analysed in this study, the {\textsc SUBFIND} algorithm \citep{springel01,dolag09} 
was used to identify galaxies.
In cases of an overlapping galaxy pair, {\textsc SUBFIND} will 
often assign some of the stellar mass from the outskirts of the lower mass galaxy to the higher mass galaxy \citep{rodriguez15}.
This effect is often referred to as numerical stripping, and it can result in spurious mass estimates for close galaxy pairs.  
Moreover, galaxies which are close to our selected lower limit in stellar mass may be erroneously removed from 
the sample by numerical stripping.  

While numerical stripping is largely irrelevant for most galaxies, it is particularly problematic 
for galaxies that are undergoing close encounters - a population that is of particular interest in our study.
We therefore attempt to minimise the effects of numerical stripping on our mass-limited samples 
by identifying systems which are affected by crowding and, in some cases, by correcting for this effect.

To begin, we extract two estimates of stellar mass for each galaxy: the stellar mass at the epoch of interest (hereafter $M_*^{\rm now}$) 
and the maximum recent stellar mass (hereafter $M_*^{\rm max}$), including the epoch of interest.
In cases where numerical stripping has occurred, $M_*^{\rm max}$ may provide an improved estimate of 
the galaxy's true current mass.  
However, if one probes too far back in time, this approach can overestimate the current stellar mass 
if there has subsequently been physical stripping (e.g. the removal of stellar mass upon infall 
into a galaxy group or cluster).

In order to minimise the effects of numerical (but not physical) stripping, we elect to evaluate $M_*^{\rm max}$ within the 
past 0.5 Gyr\footnote{Our approach is similar to that of \citet{rodriguez15}, who estimate the mass of both galaxies at the time that 
the secondary galaxy had its maximum progenitor mass.  However, as they do not apply a restriction on lookback time, 
their approach avoids both numerical and physical stripping.}.
Merger simulations suggest that this time frame should be long enough to catch the galaxies when they were  
clearly separated from one another \citep[e.g.][]{patton13}.  In addition, this time frame should be 
short enough that any physical stripping of stellar mass would be limited.  Given the time sampling 
of the simulations under consideration (see Table~\ref{tabsim}), we trace each galaxy's main progenitor branch back in time 
by 2-3 snapshots for IllustrisTNG, 2-4 snapshots for Illustris-1, and 3-5 snipshots for EAGLE.
We note also that the main conclusions of this study are not sensitive to the precise time frame used for these calculations.  
For example, changing the time frame by $+/-$ 0.25 Gyr has a negligible impact on our results.

We then quantify the degree of crowding in each galaxy's immediate surroundings by measuring the 
relative separation between the galaxy and its most overlapping companion, taking into account 
the separation and size of the galaxy and its companions.  
In order to ensure that we do not miss any galaxies above our $10^9 \msun$ threshold, 
we initially consider all galaxies with $M_*^{\rm max} > 10^9 \msun$ (thereby including some galaxies 
which have $M_*^{\rm now} < 10^9 \msun$), and we search for all companions with $M_*^{\rm max}$ 
greater than 10 per cent of the host galaxy's $M_*^{\rm now}$.
We calculate the relative separation (hereafter $\rsep$) of each neighbouring galaxy, 
as follows: 
\begin{equation}
\rsep = {r \over \rhalfh+\rhalfc},  
\end{equation}
where $r$ is the 3D separation between the centres of the two galaxies 
and $\rhalfh$ and $\rhalfc$ are the stellar half mass radii of the host and companion respectively.
For each host galaxy, we record the minimum $\rsep$ of its neighbouring galaxies.

\begin{figure}
\centerline{\rotatebox{0}{\resizebox{9.0cm}{!}
{\includegraphics{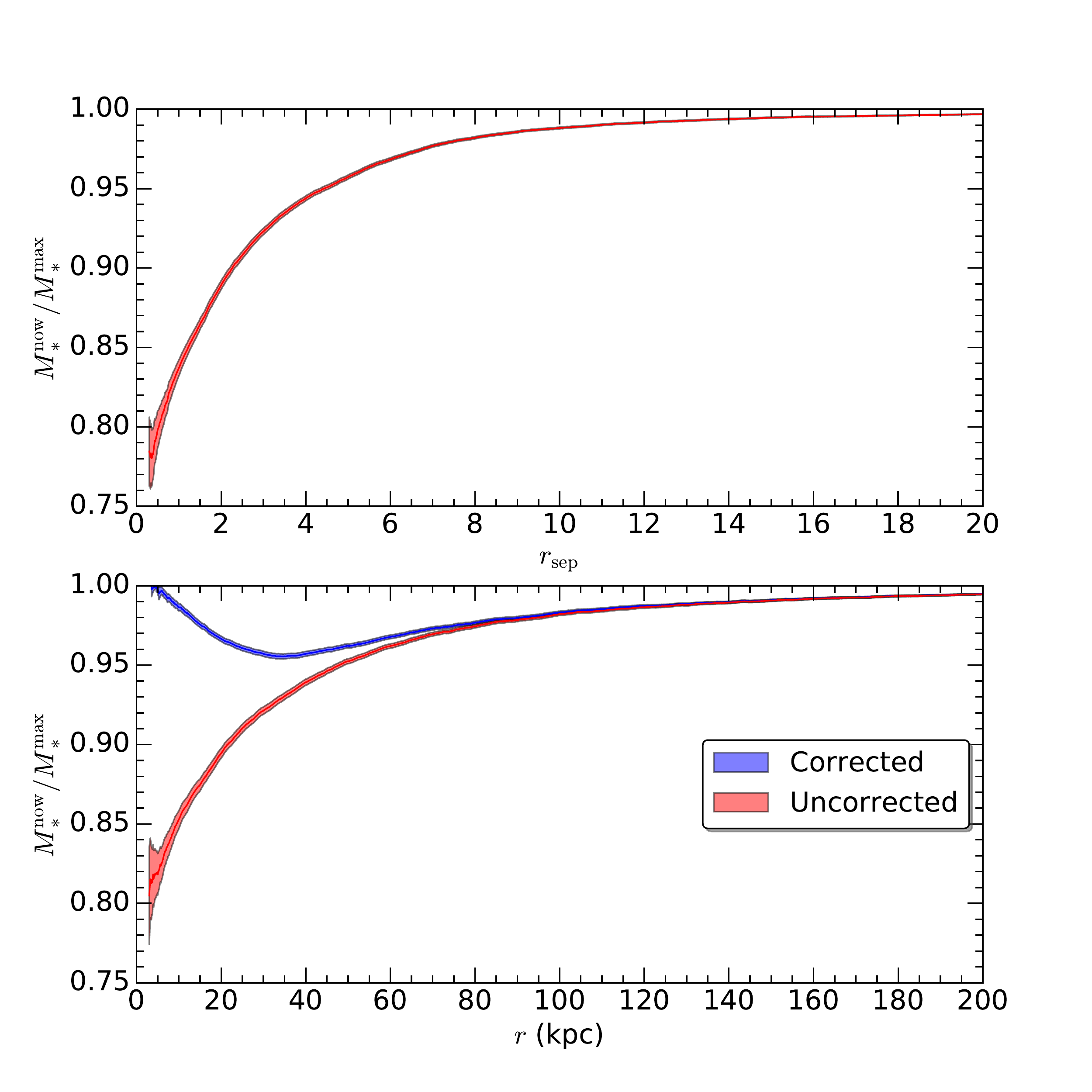}}}}
\caption{\new{In the upper panel, the mean ratio of the current stellar mass ($M_*^{\rm max}$) to the maximum stellar mass 
within the past 500 Myr ($M_*^{\rm max}$) is plotted versus the relative separation ($r_{\rm sep}$) between each galaxy 
and its closest companion in TNG100-1.  
In the lower panel, mean $M_*^{\rm now}/M_*^{\rm max}$ is plotted versus the 3D distance ($r$) between each galaxy 
and its closest companion for both the ``Uncorrected'' sample (using the true $M_*^{\rm now}$ in each case) 
and the ``Corrected'' sample (for which we replace $M_*^{\rm now}$ with $M_*^{\rm max}$ when $r_{\rm sep} < 2$).  
For each set of data, the solid line depicts the mean, and the shaded region depicts the 2$\sigma$ standard 
error in the mean.}\label{figmnowmmax}}
\end{figure}

\new{We investigate the dependence of stripping on the relative separation between TNG100-1 galaxies 
by plotting the mean $M_*^{\rm now}/M_*^{\rm max}$ versus $r_{\rm sep}$ in the upper panel of Fig.~\ref{figmnowmmax}.
At large $r_{\rm sep}$, $M_*^{\rm now}/M_*^{\rm max}$ is close to unity, indicating that few galaxies 
have experienced a recent decline in their stellar mass.  This ratio initially declines gradually as $r_{\rm sep}$ decreases, 
reaching 90 per cent at $r_{\rm sep} \sim 2$.  Given that galaxies are relatively well separated at $r_{\rm sep} > 2$, we 
interpret this trend as being driven primarily by physical stripping.
$M_*^{\rm now}/M_*^{\rm max}$ declines sharply at smaller $r_{\rm sep}$, 
dropping below 80 per cent at $r_{\rm sep} \sim 0.5$.  Given that there is substantial overlap between galaxies at these 
small relative separations, we attribute this more rapid decline as being caused by numerical stripping.}

In order to \new{identify the regime in which numerical stripping is prevalent}, we carried out an extensive visual inspection 
of synthetic stellar images of galaxies from both TNG100-1 \citep{nelson19} and Illustris-1 \citep{torrey15}.
\new{This inspection focussed on systems with $\rsep < 3$, which corresponds to typical 
3D separations of $r \lesssim 30$ kpc for TNG100-1 and $r \lesssim 60$ kpc for Illustris-1.}
At $\rsep < 1$ (i.e. for galaxy pairs with largely overlapping stellar distributions), 
we find that \new{nearly every system appears to suffer} from substantial numerical stripping, with a significant fraction 
of a galaxy's stellar mass being assigned to a close neighbour (or vice versa).  
For these galaxies, we therefore set $M_* = M_*^{\rm max}$.  
Conversely, at $\rsep > 2$, the  {\textsc SUBFIND}  algorithm \new{appears to assigns the stellar mass to the correct galaxy 
in almost every instance.   
We therefore set $M_* = M_*^{\rm now}$ when $\rsep > 2$}.

\begin{figure}
\centerline{\rotatebox{0}{\resizebox{9.0cm}{!}
{\includegraphics{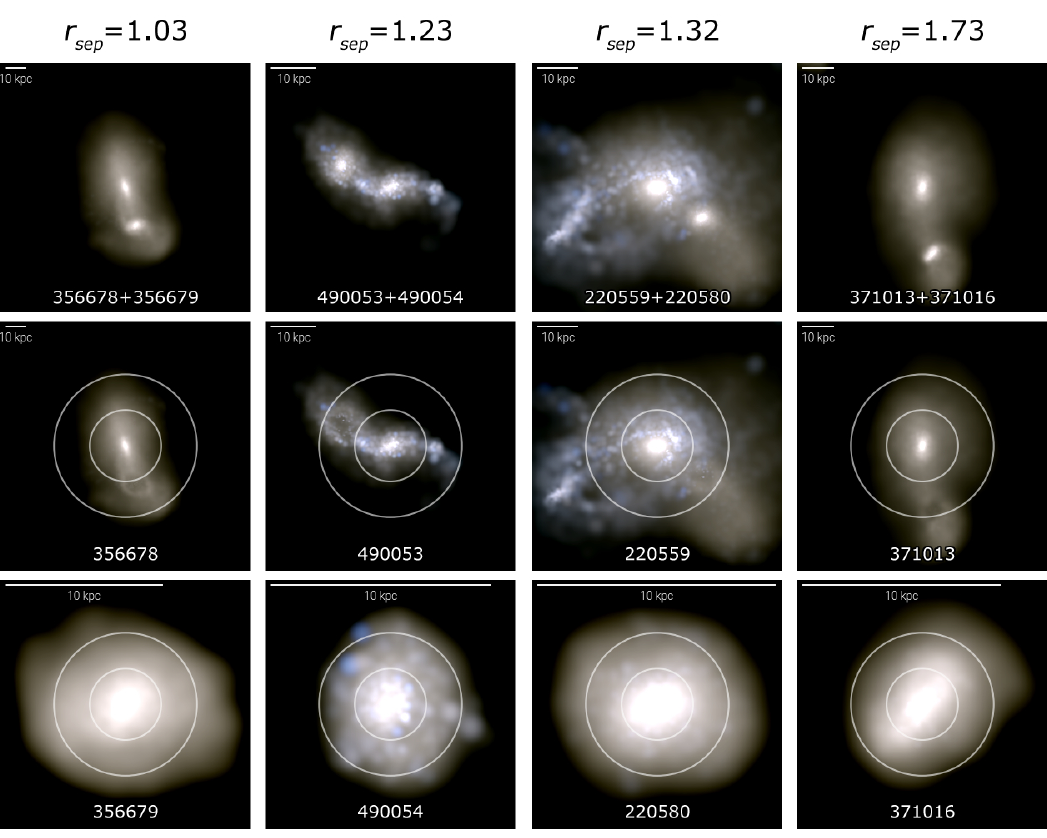}}}}
\caption{Synthetic stellar composite images of four $z$=$0$ TNG100-1 host galaxies and their closest companions are shown, 
with one column corresponding to each pair.  The relative separation of each pair is shown at the top.
In the upper row, each image is centred on the host galaxy and shows all of the subhaloes within the specified volume (i.e., both the host and companion are shown).  
In the middle row, each image is again centred on the host galaxy, but in this case only the stellar mass associated with the host galaxy is shown.  
In the lower row, each image is centred on the companion galaxy, and only the stellar mass associated with the companion galaxy is shown.  
In the middle and lower rows, white circles are used to denote one and two $\rhalf$.
All images are labelled with the galaxy SubfindIDs.  
The width of each image corresponds to 7 $\rhalf$ of the galaxy at the centre of the image, 
and the white bar at the upper left of each image depicts a physical scale of 10 kpc.
\label{figrsepim}}
\end{figure}

Within the intermediate range of $1 < \rsep < 2$,  {\textsc SUBFIND} tends to do a reasonable but sub-optimal job 
at separating the two galaxies.  We illustrate this in Fig.~\ref{figrsepim}, 
where we present stellar composite images\footnote{These images were generated using the IllustrisTNG 
image visualisation tool 
available at http://www.tng-project.org, with the Stellar Composite field.  See \citet{nelson19} for details.} 
of four TNG100-1 galaxies that have a close companion at $1 < \rsep < 2$.  
\new{For each of these galaxy pairs, we show images of the host plus companion (top row), host only (middle row), 
and companion only (bottom row).}
These pairs are oriented such that their projected separation is at least 90 per cent of 
their 3D separation, making the images easier to interpret.
In each case, we can clearly see that some stellar mass 
is incorrectly assigned to the more massive member of the pair.  
To be conservative, we therefore set $M_* = M_*^{\rm max}$ for all galaxies with $\rsep < 2$.  

\new{In order to assess the impact of this procedure on our sample, we plot the mean $M_*^{\rm now}/M_*^{\rm max}$ 
versus the 3D distance to the closest companion ($r$) in the lower panel of Fig.~\ref{figmnowmmax}.
For the original ``Uncorrected'' TNG100-1 sample, $M_*^{\rm now}/M_*^{\rm max}$ drops below 90 per cent 
at $r \lesssim 20$ kpc, presumably driven by numerical stripping.  However, after replacing 
$M_*^{\rm now}$ with $M_*^{\rm max}$ for $\rsep < 2$, the ``Corrected'' mean $M_*^{\rm now}/M_*^{\rm max}$ 
remains above 95 per cent at all separations, thereby compensating for numerical stripping.
This procedure increases} the stellar mass for only 1.2 per cent of the galaxies 
in TNG100-1, confirming that only a small fraction of the sample is significantly affected by numerical stripping.
With our best estimates of $M_*$ in hand for all of the galaxies in each simulation, we then create 
our desired mass limited samples by requiring $M_* > 10^9 \msun$.  

\subsection{Host galaxies and their closest companions}

The primary goal of this study is to identify the statistical influence (if any) of 
closest companions on galaxy sSFRs.  
We follow the methodology of \citet{patton13} and \citet{patton16} to identify the
closest companions of massive galaxies (hereafter host galaxies) in the simulations.
\new{While we do not attempt to detect the influence of additional companions, 
earlier studies have shown that more distant companions are likely to 
play a relatively minor role \citep{moreno13,an19}.}

We restrict our analysis to host galaxies in the range $10^{10}\msun < M_* < 10^{12}\msun$.  
The minimum mass is a factor of ten larger than the limiting stellar mass of our sample, 
ensuring adequate depth when searching for companions.
The maximum mass of $10^{12} \msun$ excludes the high mass tail of the stellar mass distribution,  
which contains galaxies which are more susceptible to biases due to overlapping subhaloes (see 
Section~\ref{secmass} above).

For each host galaxy, we find the closest companion, using 3D physical separation (hereafter $r$) to quantify proximity.  
We require all potential companions to have a companion stellar mass ratio (hereafter $\mu_{\rm c}$, 
where $\mu_{\rm c} = M_*^{\rm comp}/M_*^{\rm host}$) of at least 0.1. 
Given that host galaxies have $M_* > 10^{10}\msun$, all potential companions will lie above the stellar mass limit 
of our sample ($10^{9}\msun$).  We note that companions may be more or less massive than their host galaxy.

In addition to identifying each galaxy's closest companion, we also characterise each galaxy's environment 
using the 3D equivalent of the $N_2$ and $r_2$ parameters of \citet{patton16}.  
$N_2$  refers to the number of companions ($\mu_{\rm c} \ge 0.1$) that lie within 2 Mpc of the host galaxy, and $r_2$ refers to 
the distance to the host galaxy's second closest companion.   The former is related to the local density of galaxies, 
while the latter quantifies the isolation of a galaxy pair from its surroundings.   
\new{For example, a galaxy with small $r$ and large $r_2$ is a member of an isolated pair, whereas a galaxy with 
small $r$, small $r_2$ and large $N_2$ is likely to be located in a rich group or cluster \citep{patton16}.}
We also recompute each galaxy's $\rsep$ at this stage, using our best estimates of stellar mass ($M_*$) rather than 
$M_*^{\rm now}$ and $M_*^{\rm max}$ (see Sec.~\ref{secmass}).

Finally, in order to focus our analysis on interactions between galaxies with masses that are within a factor of 10 of one another, 
we also exclude host galaxies whose closest companion has $\mu_{\rm c} > 10$.   This restriction removes low mass host galaxies which are 
in the vicinity of a massive galaxy.   \new{For all of the simulations under consideration, we find similar numbers of galaxies in 
major versus minor pairs.  For example, 47 per cent of TNG100-1 galaxies have a closest companion with $0.3 \le \mu_{\rm c} \le 3.\overline{3}$.
We also find that most galaxies are more massive than their closest companion (e.g. this is true for 77 per cent of galaxies in TNG100-1).
This is to be expected, given that the number density of galaxies - and hence potential companions - decreases with increasing stellar mass.}

\subsection{sSFR estimates}\label{secssfr}

All of the hydrodynamical simulations used in this study implement a sub-resolution model for star formation 
that occurs stochastically within dense interstellar gas.
The SFRs of the simulated galaxies are computed by summing the instantaneous SFRs of all gas cells/particles bound to the subhalo.  

For our analysis, we elect to consider only the star formation that occurs within the 
central stellar half mass radius ($\rhalf$) of each galaxy.  This choice enables us to avoid the outskirts of galaxies, where 
there is an increased chance of including star-forming gas from a galaxy's surroundings (especially when there are 
neighbouring galaxies).  Moreover, galaxy pair observations indicate that star formation enhancements 
tend to be centrally concentrated \citep{ellison13,barrera15,chown19,thorp19,pan19}, as predicted by merger simulations 
\citep{mihos94,cox06,dimatteo07,moreno15}.
This suggests that we are likely to capture the majority of the enhanced star formation within $\rhalf$.

\new{For each galaxy, we compute the sSFR by dividing the SFR within $\rhalf$ by 
the stellar mass within $\rhalf$.}
These quantities are provided in the IllustrisTNG and Illustris databases. 
The EAGLE database instead reports the SFR and the stellar mass within a range of apertures \citep{mcalpine16}.  
We therefore interpolate between these apertures to estimate the sSFR within $\rhalf$ for EAGLE galaxies.  

For the majority of host galaxies, this approach yields a secure estimate of the sSFR within the 
central $\rhalf$.  However, in the cases where the host galaxy has a close companion, it is possible that 
the incorrect assignment of particles by the {\textsc SUBFIND} algorithm may lead to an unreliable estimate of the 
host galaxy's sSFR.  Our primary concern is to avoid situations in which the stellar mass and/or star forming gas 
from a companion contaminates the host galaxy's sSFR.  Our secondary concern is to avoid cases where 
the host galaxy's $\rhalf$ is significantly underestimated or overestimated as a result of overlap with the companion.

We have previously addressed this type of issue when assigning total stellar masses to galaxies in Section~\ref{secmass}.  
In the case of measuring sSFR within $\rhalf$, the problem is certain to be less severe, since sSFR is a relative 
(rather than absolute) quantity, and because numerical stripping preferentially affects the outer regions of galaxies.
Having visually inspected synthetic images of approximately 100 randomly-selected IllustrisTNG and Illustris-1 host galaxies with close companions, 
we conclude that sSFR measurements are likely to be reliable for galaxies whose closest companion lies at $\rsep>1$.  
For example, Fig.~\ref{figrsepim} shows the locations of $\rhalf$ for four TNG100-1 pairs 
with $1<\rsep<2$.  In each case, the region within the host galaxy's $\rhalf$ appears to suffer from little (if any) 
contamination from the companion. 

In cases where the closest companion lies at $\rsep<1$, it becomes very difficult to accurately separate the two galaxies.  
This makes the sSFR estimates quite uncertain.  We therefore elect to remove from our analysis all host galaxies 
which have $\rsep < 1$.  In Table~\ref{tabsim}, we report the average number of host galaxies per snapshot after 
this restriction has been applied to all of the simulations under consideration. 
In the following section, we address the incompleteness in our samples that results from this restriction.  

\subsection{Crowding analysis}\label{seccrowd}

We have excluded from our analysis all host galaxies whose closest companion lies at $\rsep < 1$ (see Section~\ref{secmass}).   
While this will help to ensure that we avoid galaxies whose sSFR estimates are unreliable due to crowding, 
we also wish to avoid separations at which a substantial fraction of the sample has been excluded because of crowding.
Our primary concern is that the largest host galaxies are the most likely to overlap with their closest companion 
and be removed from our sample, and we wish to avoid biasing our sample towards smaller galaxies at small separations.

\begin{figure}
\centerline{\rotatebox{0}{\resizebox{9.0cm}{!}
{\includegraphics{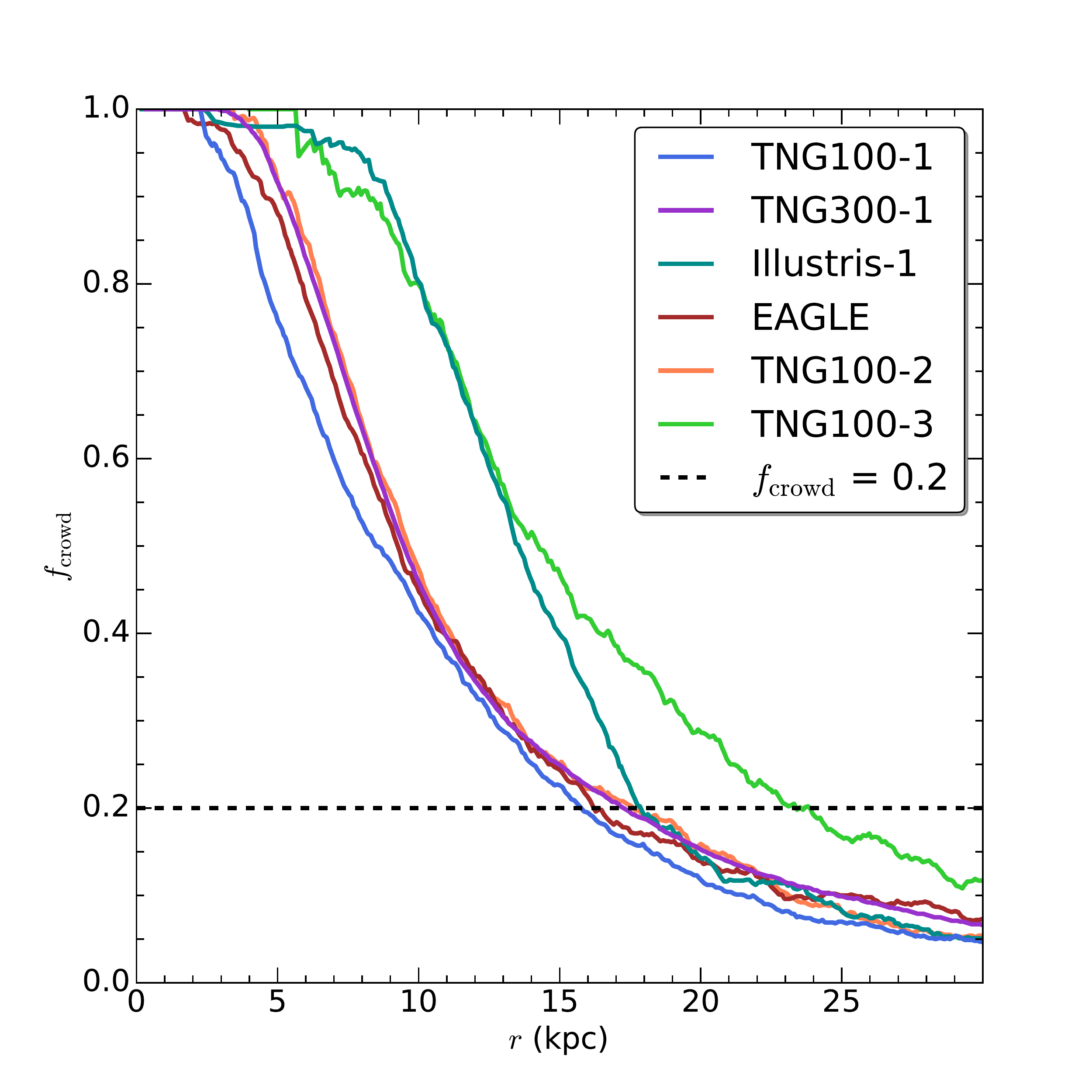}}}}
\caption{The crowding fraction $f_{\rm crowd}$ is plotted versus the 3D distance to the closest companion ($r$) 
for all of the simulations used in this study.  The horizontal dashed line depicts a crowding fraction of 20 per cent.
\label{figrsep}}
\end{figure}

In order to identify the separation below which crowding leads to significant incompleteness, we compute the crowding fraction (hereafter $f_{\rm crowd}$) 
as a function of the 3D separation $r$, where $f_{\rm crowd}$ is the fraction of galaxies that have been removed from 
our sample by the $\rsep > 1$ requirement.  In Fig.~\ref{figrsep}, we plot $f_{\rm crowd}$ vs. $r$ for all of the simulations 
used in this study.  This figure demonstrates that most of the simulations exhibit high crowding fractions at $r \lesssim 10$ kpc, 
and all but one (TNG100-3) have crowding fractions that drop below 10 percent above 25 kpc.  The TNG100-2 and TNG300-1 simulations, 
which have the same resolution but very different volumes, have nearly identical crowding fractions.  
Comparing the three TNG100 simulations with different resolutions, we see a clear increase in crowding as 
we move from high resolution (TNG100-1) to low resolution (TNG100-3).  

In addition, if we compare the three simulations with similar resolution and volume, we detect significant differences 
in crowding, with the TNG100-1 simulation being less susceptible to crowding than EAGLE and especially Illustris-1.
This disparity likely arises from differences in the simulation subgrid model prescriptions 
(particularly their differing implementations of stellar feedback) and calibration strategy, which lead to differences in galaxy sizes.
By comparing the mean sizes of galaxies in these three simulations, 
we find that TNG100-1 galaxies are somewhat smaller than those in EAGLE, while Illustris-1 galaxies are substantially larger.   
This is consistent with published analyses of galaxy sizes in these simulations \citep{snyder15,bottrell17,furlong17,genel18,rodriguez19}, 
and suggests that galaxy sizes may be driving these differences in crowding.

Having characterised the degree of crowding in each of the simulations, we wish to apply a consistent crowding threshold that 
will ensure that all of the simulations we use are largely unaffected by crowding.  In doing so, we also 
wish to retain sizeable samples of galaxies with relatively close companions.  
In the analysis that follows, we use $f_{\rm crowd}$ = 0.2 as an acceptable compromise, 
and we compute the separation $r_{\rm crowd}$ at which $f_{\rm crowd}$ = 0.2 for each of the simulations.  
On many of the figures that follow, we identify regions where the crowding fraction exceeds this threshold (i.e., where $r < r_{\rm crowd}$), 
and we avoid analysing sSFR values at these small separations.

\subsection{Combining samples from a range of redshifts}

Following the procedures described above, we create samples of massive host galaxies  
with reliable estimates of stellar mass and sSFR at every available snapshot from $z\sim1$ to $z=0$.   
As summarised in Table~\ref{tabsim}, this yields an average of $\sim$ 2500-6000 host galaxies per snapshot 
for each of the $L_{\rm box}\sim100$ Mpc simulations under consideration.  
These samples are too small to provide tight constraints on sSFR enhancements as a function of closest companion separation at a given epoch, 
given that only a few percent of galaxies have close companions at low redshift \citep{patton00,patton08,robotham14}. 

However, we have created samples of host galaxies at $\sim$ 50 different snapshots for each of the IllustrisTNG (and Illustris) simulations.
These snapshots are separated in time by an average of about 160 Myr, during which time any pair of interacting galaxies 
can undergo meaningful changes in separation, orientation, sSFR and (to a lesser extent) stellar mass.  
As such, it is reasonable to consider each snapshot as a semi-independent set of data from which we can learn about the 
influence of close companions on their host galaxies.  Therefore, for each simulation, we combine all of our host galaxy catalogues 
at $z < 1$, yielding much larger statistical samples than we have at any single epoch.   For example, 
this yields a sample of nearly 300,000 host galaxies in TNG100-1 and roughly 4.5 million host galaxies in TNG300-1.
We later investigate how our results vary with redshift within these samples (see Section~\ref{secrssfrz}).

\subsection{Creation of statistical control samples}

A key goal of this study is to assess the sSFR enhancement in galaxies 
that is associated with the presence of a close companion.  
For merger simulations, sSFR enhancements can be estimated by 
tracking changes in a galaxy's sSFR as it proceeds through the merger sequence.
To account for sSFR changes that would have occurred in the absence of an interaction, 
these enhancements are often computed by comparing the merging galaxies 
with identical galaxies that are evolved 
in isolation \citep{cox08,torrey12,scudder12,patton13,moreno19}. 

This approach accurately quantifies the additional star formation that is triggered by 
an interaction within a simulation.  However, while we could in principle track changes 
in sSFR over time for galaxies in our cosmological simulations, we do not know 
how they would have evolved in isolation.  Moreover, observers are obviously never able 
to quantify sSFR enhancement using this approach.

We instead elect to use control samples to assess enhancements in sSFR, 
following the methodology used by \citet{patton16} for their observational sample.
This approach has the added benefit of facilitating a more direct comparison between 
enhanced star formation in cosmological simulations and observations (see Section~\ref{secsdss}).

We now describe the creation of our control sample.
For each host galaxy, we identify control galaxies which are closely matched to each host galaxy's 
redshift, stellar mass, local density, and isolation.  For the latter, we require that the distance to the 
control galaxy's closest companion ($r$) be closely matched to the distance to the host galaxy's 
{\it second closest} companion ($r_2$).   This criterion ensures that the host galaxy pair 
and the control galaxy are similarly isolated from their surroundings.
\new{When combined with matching on local density, this approach ensures that host galaxies 
and their controls reside in similar environments}, thereby 
separating out the influence of the host galaxy's closest companion.   

Given that the number of host galaxies at a given epoch is relatively small in most of the simulations 
under consideration (see Table~\ref{tabsim}), it is challenging to find numerous closely matched controls 
for each host galaxy; we therefore elect to identify the single best control for each host galaxy.
We allow a given galaxy to be a host as well as the control for 
one or more other galaxies.

To match on redshift, we require an exact match at the time of the simulation output (i.e. snapshot), 
thereby ensuring that our host and control samples will have identical redshift distributions.
For stellar mass, we begin with a matching tolerance of 0.05 dex.  
For local density, we initially require $N_2$ to match within 10 per cent.  
For isolation, we initially require the host galaxy's $r_2$
to agree with its control galaxy's $r$ within 10 per cent.  
To ensure a fair distribution of controls within all of these tolerances, 
we further narrow our sample of host galaxies to those with $N_2 \ge 2$, 
$r_2 \le 1500$ kpc and $10.05 \le \log (M_*/\msun) \le 11.95$.

In most cases, these initial tolerances yield at least one control galaxy for every host galaxy.  In cases where no controls are identified, 
we increase all three of the tolerances by 50 per cent and redo the search, repeating once more if no controls are found.
Finally, if more than one control has been found, we choose the control with the best simultaneous match 
in stellar mass, $N_2$ and $r_2$, using the weighting scheme of \citet{patton16}.  

\begin{figure}
\centerline{\rotatebox{0}{\resizebox{9.0cm}{!}
{\includegraphics{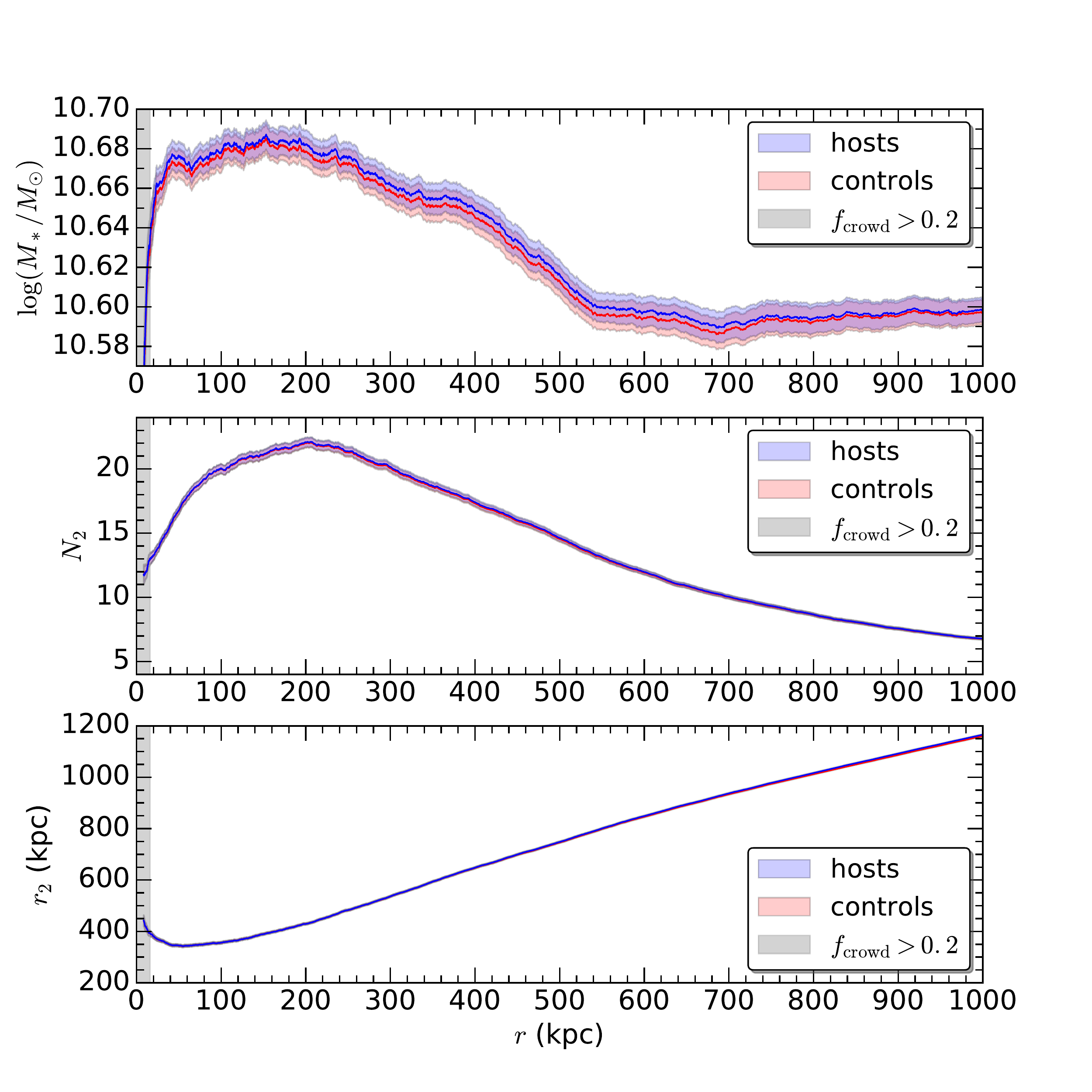}}}}
\caption{The galaxy properties which are used for control sample matching are plotted versus the 3D distance to 
the host galaxy's closest companion for the TNG100-1 simulation.  
The upper panel shows the log of the mean stellar mass ($\log(M_*/\msun)$) of host galaxies 
and their best controls.
The middle panel shows the mean local density ($N_2$) of hosts and companions.  
The lower panel shows the mean 3D distance to the second closest companion ($r_2$) for host galaxies, and the 
mean 3D distance to the closest companion ($r$) for the controls, since these properties are matched with one another.
In all panels, the coloured shading depicts the $2\sigma$ standard error in the mean, and 
the gray shaded region on the left depicts separations at which the crowding fraction exceeds 20 per cent.
The quality of the matching is excellent, making it difficult to distinguish between hosts and their controls.
\label{figcon}}
\end{figure}

The outcome of this matching procedure is illustrated in Fig.~\ref{figcon} for galaxies in the TNG100-1 simulation.
In this figure, we plot $\log(M_*/\msun)$, $N_2$ and $r_2$ versus $r$ for host galaxies and their best controls.
The control matching is excellent; on average, TNG100-1 host galaxies and their controls agree within 6 per cent in $M_*$, 
3 per cent in $N_2$ and 5 per cent in $r_2$.   
The quality of the control matching is similar for the other $L_{\rm box}\sim100$ Mpc 
simulations used in this study, and is even better in the TNG300-1 simulation 
(which has many more potential controls to choose from for each host galaxy).

We note that while host galaxies and their controls are in close agreement at all separations, 
$\log(M_*/\msun)$, $N_2$ and $r_2$ all vary with $r$ in Fig.~\ref{figcon}.  
There is a gradual increase in mean stellar mass from $r \sim 600$ kpc to $r \sim 200$ kpc, 
although the size of this effect is small ($<$ 0.1 dex).  There is 
a sharper drop in stellar mass below $r \sim 40$ kpc (also $<$ 0.1 dex), an effect that is likely due to crowding, 
which preferentially removes larger, higher mass galaxies from the sample at small separations (see Section~\ref{seccrowd}). 
The mean local density ($N_2$) rises by a factor of about 3 from 1000 kpc to about 200 kpc, 
and then declines by about a factor of two at the smallest separations.
Finally, isolation ($r_2$) generally increases as $r$ increases (given that $r_2 > r$, this is to be expected).
All of these trends can be interpreted as environmental effects \citep{patton16}.
Most significantly, the close agreement between host galaxies and their controls ensures that these trends are unlikely 
to have a significant influence on the main results of this study. 

\section{Enhanced Star Formation in the TNG100-1 and TNG300-1 Simulations}\label{sec3d}

The primary goal of this study is to investigate the relationship between a galaxy's sSFR and the presence of a close companion.
We begin by analysing the relationship between sSFR and the 3D distance to the closest companion 
for all of the simulations introduced in Section~\ref{secmethods}, starting with the fiducial TNG model, TNG100-1.

\subsection{Enhanced sSFR vs. 3D separation in TNG100-1}\label{sect1}

\begin{figure}
\centerline{\rotatebox{0}{\resizebox{9.0cm}{!}
{\includegraphics{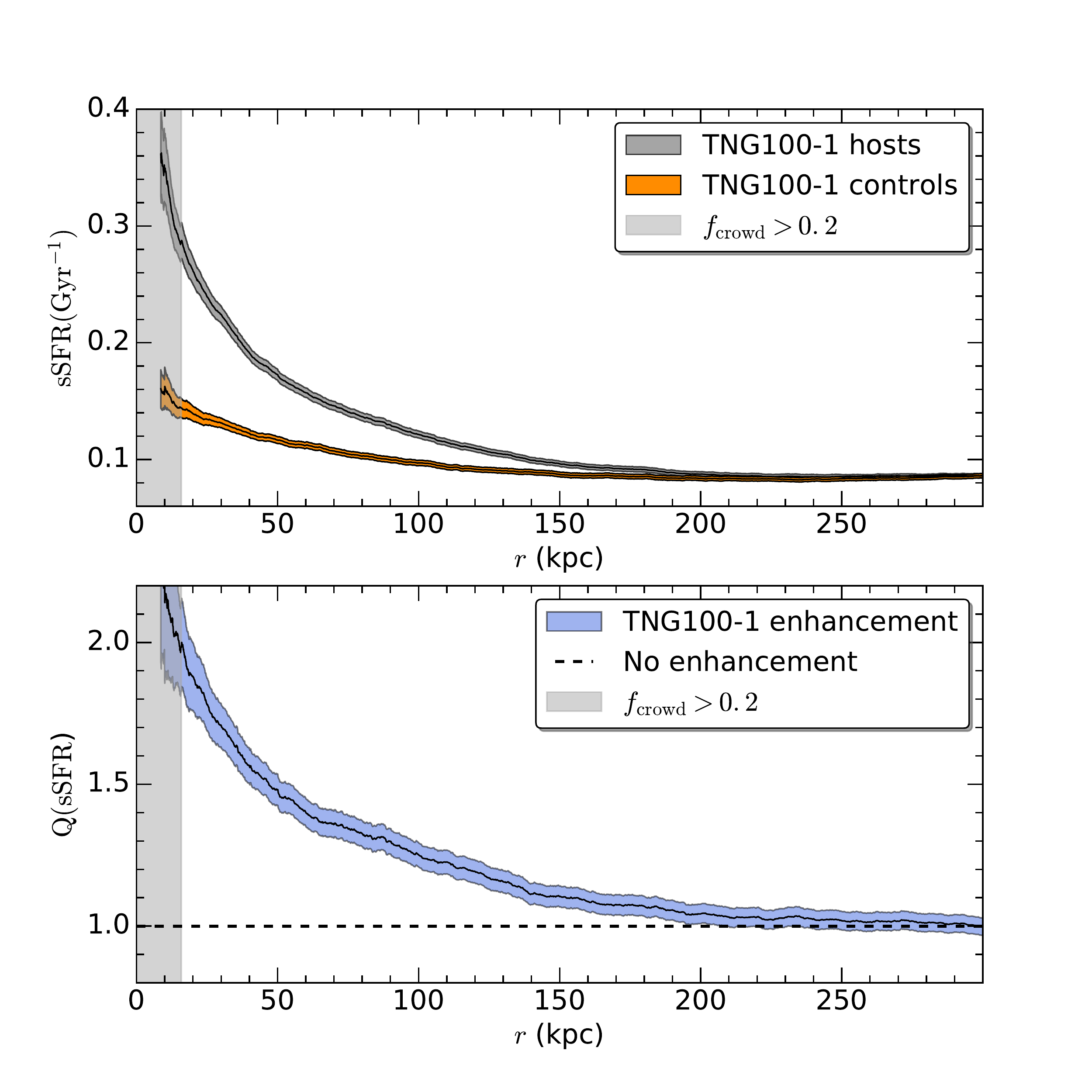}}}}
\caption{The mean sSFR of TNG100-1 host galaxies and their controls 
are plotted vs. the distance to the host galaxy's closest companion ($r$) in the upper panel.  
The mean sSFR values have been smoothed using a variable width box kernel.
The corresponding sSFR enhancements ($Q$) are plotted in the lower panel, with $Q = 1$ (dashed line) corresponding to no enhancement.
The shaded region surrounding each line shows the 2$\sigma$ standard error in the mean. 
The shaded region on the left side of both panels depicts separations 
at which $f_{\rm crowd} > 0.2$ (see Sec.~\ref{seccrowd}).
\label{figt1}}
\end{figure}

In the upper panel of Fig.~\ref{figt1}, we plot the mean sSFR of galaxies in TNG100-1 
as a function of the 3D distance to the host galaxy's closest companion ($r$).  
To provide the most useful estimate of the mean sSFR at every $r$, we smooth the 
data using a variable width box kernel\footnote{The default bin width is $\pm$ 25 kpc. 
At small separations, the bin width is decreased to match the 
range of $r$ for which data is available, while at $r > 100$ kpc, the bin width is 
gradually increased to ensure an adequate number of galaxies in each bin.
However, the exact choice of bin width does not impact our results.}.
We detect a clear increase in the mean sSFR as $r$ decreases, with the mean 
sSFR increasing by a factor of $\sim$ 3 from 0.09 Gyr$^{-1}$ at $r \sim 200$ kpc 
to 0.29 Gyr$^{-1}$ at the crowding limit of 15.9 kpc.  

In the upper panel of Fig.~\ref{figt1}, we also plot the mean sSFR of the best controls for these galaxies
as a function of the 3D distance to the corresponding {\it host} galaxy's closest companion.  
Beyond 3D separations of about 200 kpc, we find no statistically significant difference between 
the mean sSFR of the host galaxies and their controls.  At smaller separations, however, there is 
a clear separation in the trends of host galaxies and their controls, with a much smaller rise 
in the mean sSFR of the controls as $r$ decreases.  The modest rise in the 
mean sSFR of the controls at small $r$ is likely due to the fact that the controls are 
not strictly isolated; rather, the closest companions of the controls are matched 
to the separation of the second closest companions of their host galaxies.  
That is, our methodology is designed to detect the influence 
of the closest companion, but it does not capture the influence of any additional 
companions\footnote{Most of the rise in the mean sSFR of the control sample at small $r$ 
disappears if we restrict the host galaxy sample to $r_2 > 100$ kpc, thereby creating a relatively isolated sample of galaxy pairs.}.  

In order to more directly compare the sSFRs of host galaxies and their controls, we 
also estimate the sSFR enhancement (hereafter $Q$) in a given $r$ bin by dividing the mean sSFR of the host galaxies 
by the mean sSFR of their controls; i.e., 
\begin{equation}\label{eqnQ}
Q(\mathrm{sSFR}) = {\langle \mathrm{sSFR}_{\rm hosts}\rangle \over \langle \mathrm{sSFR}_{\rm controls}\rangle}.
\end{equation}
In the lower panel of Fig.~\ref{figt1}, we see values of $Q\sim 1$ at $r \gtrsim 200$ kpc, 
indicating that there is no statistically significant enhancement of the host galaxy sSFRs at large separations.  
Conversely, below 200 kpc, sSFR enhancements increase steadily as the separation decreases,  
reaching a maximum value of $Q \sim 2.0 \pm 0.1$ (a factor of two enhancement in sSFR) at the   
crowding threshold of this sample ($r_{\rm crowd} = 15.9$ kpc).  
Moreover, this steady increase extends well into the region affected by crowding ($r < r_{\rm crowd}$), 
suggesting that our choice of $f_{\rm crowd} = 0.2$ may be quite conservative. 
Statistically significant (2$\sigma$) enhancements in the mean sSFR extend out to 210 kpc,
and have very high statistical significance ($12\sigma$) at the smallest separations we probe ($r = r_{\rm crowd}$).

\subsection{Synthetic images of TNG100-1 host galaxies with close companions}

Given the increase in mean sSFR towards small $r$ that we have found, 
it is natural to speculate that galaxies in these simulations are exhibiting enhancements in their star formation due to close encounters, 
as has been reported in high resolution idealised merger simulations and cosmological zoom-in simulations.   
One obvious test of this interpretation is to look for morphological signatures of interactions and disturbances 
in the synthetic images of these galaxies.  

\begin{figure*}
\centerline{\rotatebox{0}{\resizebox{18.0cm}{!}
{\includegraphics{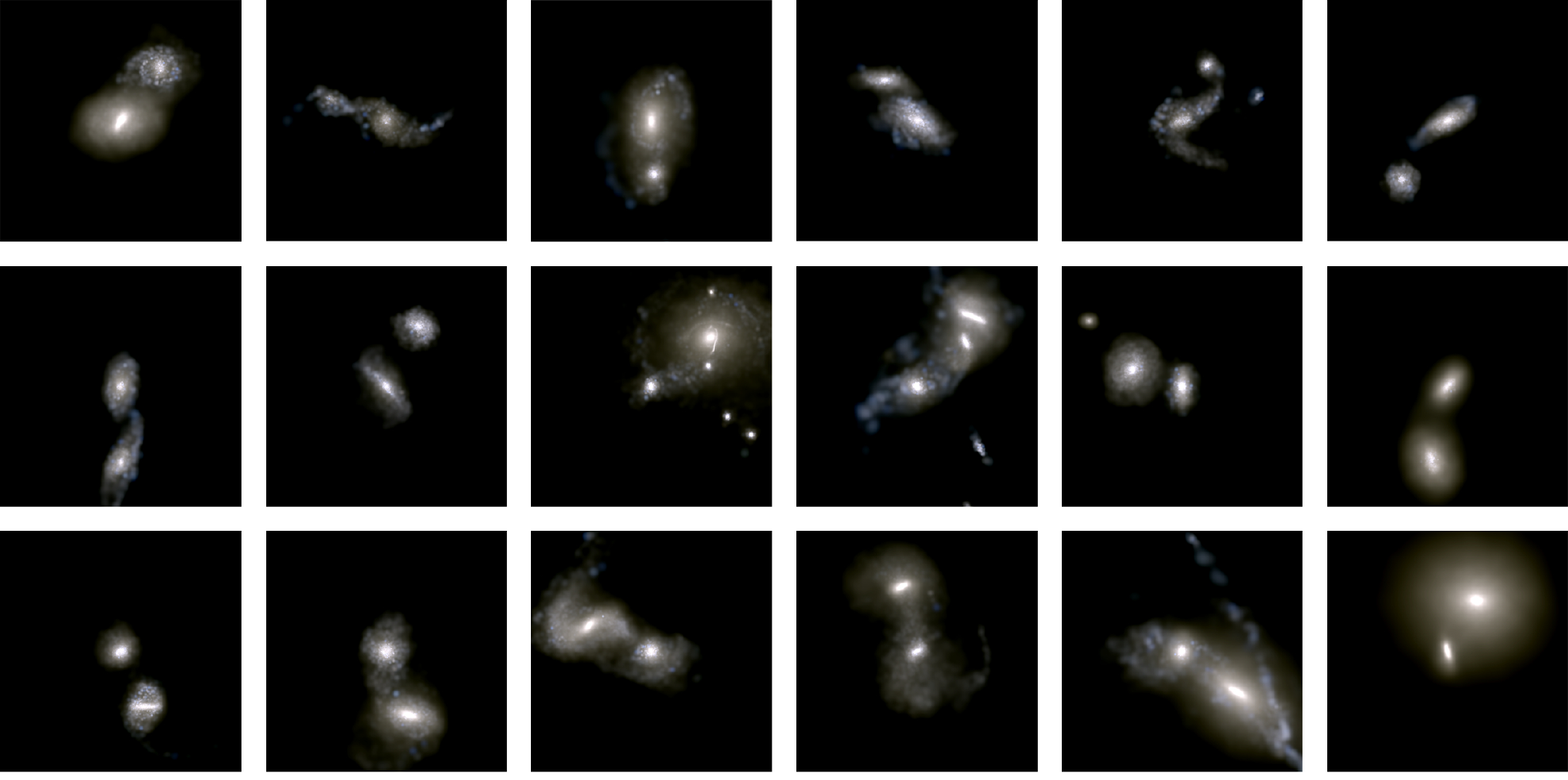}}}}
\caption{Synthetic stellar composite images of 18 TNG100-1 host galaxies with close companions.  All images are 80 physical kpc on a side, 
and are centred on the host galaxy.  
Each image depicts the host galaxy, its closest companion, and any other stellar mass in the volume that is shown.
The upper row contains host galaxies with a relatively low mass close companion (companion stellar mass ratio $0.1 < \mu_{\rm c} < 0.5$), 
the middle row contains host galaxies with a roughly equal mass close companion ($0.5 < \mu_{\rm c} < 2$), and 
the lower row contains host galaxies with a relatively high mass close companion ($2 < \mu_{\rm c} < 10$). 
\label{figt1mosaic}}
\end{figure*}

We present stellar composite images of 18 representative TNG100-1 host galaxies 
with relatively close companions ($r \sim 20$ kpc) in Fig.~\ref{figt1mosaic}.  
At these separations, the mean sSFR is enhanced by a factor of $\sim 2$ for this simulation (see Fig.~\ref{figt1}).
To cover a representative sample of galaxies in both major and minor pairs, 
this sample includes images of 6 host galaxies at low redshift ($z < 0.5$) in each of the following ranges of the companion stellar mass ratio: 
$0.1 < \mu_{\rm c} < 0.5$ (upper row), $0.5 < \mu_{\rm c} < 2$ (middle row), and $2 < \mu_{\rm c} < 10$ (lower row).
In addition, to aid the interpretation of these images, we select pairs which have projected separations (in the image plane) 
which are within 10 per cent of their 3D separations.  
Apart from these criteria, these host galaxies were selected at random from the TNG100-1 sample.
To be explicit, these galaxies were not selected based on their visual morphologies or other galaxy properties.  

Morphological signs of interactions are clearly visible in nearly all of the galaxy pairs depicted in Fig.~\ref{figt1mosaic}.  
A diversity of galaxy types and pairs is seen, including pairs of star-forming galaxies (potential wet mergers), 
pairs of passive galaxies (potential dry mergers), and star-forming+passive pairs (potential mixed mergers).   
For host galaxies which have controls with non-zero sSFRs, the median $Q$ is 2.1, 
confirming that our random selection has not generated a subset of galaxies with unusually high (or low) sSFR enhancements. 
We conclude that the enhanced sSFR we have detected in TNG100-1 does indeed appear 
to be associated with interactions between galaxies, at least at the relatively small separations ($r \sim 20$~kpc) where the sSFR enhancements are the strongest.

\subsection{Enhanced sSFR vs. 3D separation in TNG300-1}\label{sect3}

The TNG300-1 simulation allows us to investigate trends in the enhanced sSFR 
within a volume that is about 20 times larger than TNG100-1, albeit at lower resolution (see Table~\ref{tabsim}).  
This larger volume substantially reduces the statistical uncertainties on the properties measured, 
and also spans a wider range of cosmological environments.

\begin{figure}
\centerline{\rotatebox{0}{\resizebox{9.0cm}{!}
{\includegraphics{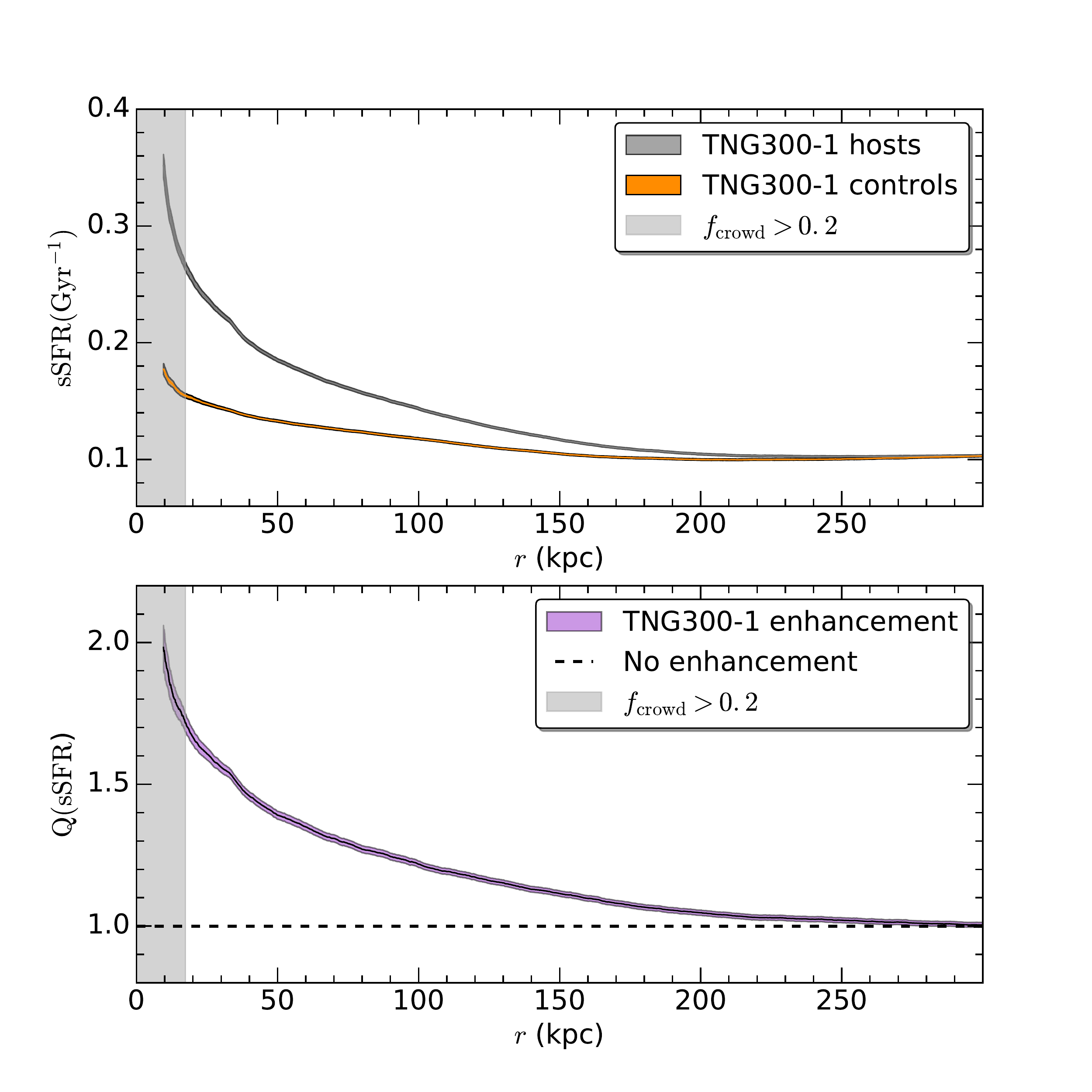}}}}
\caption{The mean sSFR of TNG300-1 host galaxies and their controls are shown in the upper panel, while  
the corresponding sSFR enhancement is shown in the lower panel.  
The shaded region surrounding each line shows the 2$\sigma$ standard error in the mean. 
\label{figt3}}
\end{figure}

In Fig.~\ref{figt3}, we show the mean sSFR and sSFR enhancement for the TNG300-1 volume.  
In both panels, we see trends that are qualitatively similar to those of TNG100-1 (Fig.~\ref{figt1}), 
yet yielding much smaller uncertainties.  Close inspection of this figure and the underlying data reveal that 
the mean sSFR is enhanced ($Q > 1$) at the $2\sigma$ level for $r < 280$ kpc.
While we restrict Fig.~\ref{figt3} to $r < 300$~kpc, we have also extended the analysis out to 1000 kpc (not shown), 
finding that $Q$ is consistent with 1.0 (within 2$\sigma$) throughout the range 300-1000 kpc for TNG300-1.
The enhancements increase steadily as $r$ decreases, reaching $Q = 1.72 \pm 0.02$ at the smallest separation for which 
$f_{\rm crowd} < 0.2$ (17.4 kpc)\footnote{The enhancements reach $Q = 1.98 \pm 0.04$ at 
the smallest separation shown (9.5 kpc), which corresponds to $f_{\rm crowd}$ = 0.5 (at that separation, half of 
the host galaxies have been excluded due to crowding).}.

\subsection{Redshift dependence}\label{secrssfrz}

As mentioned previously, our sample includes host galaxies in the redshift range $0 \leq z < 1$.  
The sSFR enhancements we have reported should be representative of galaxies throughout this 
redshift range, as long as there is no strong redshift evolution in the mean sSFR enhancement 
at $z < 1$.  We now assess the validity of this 
approach by dividing our TNG300-1 sample into five narrower redshift bins.  

\begin{figure}
\centerline{\rotatebox{0}{\resizebox{9.0cm}{!}
{\includegraphics{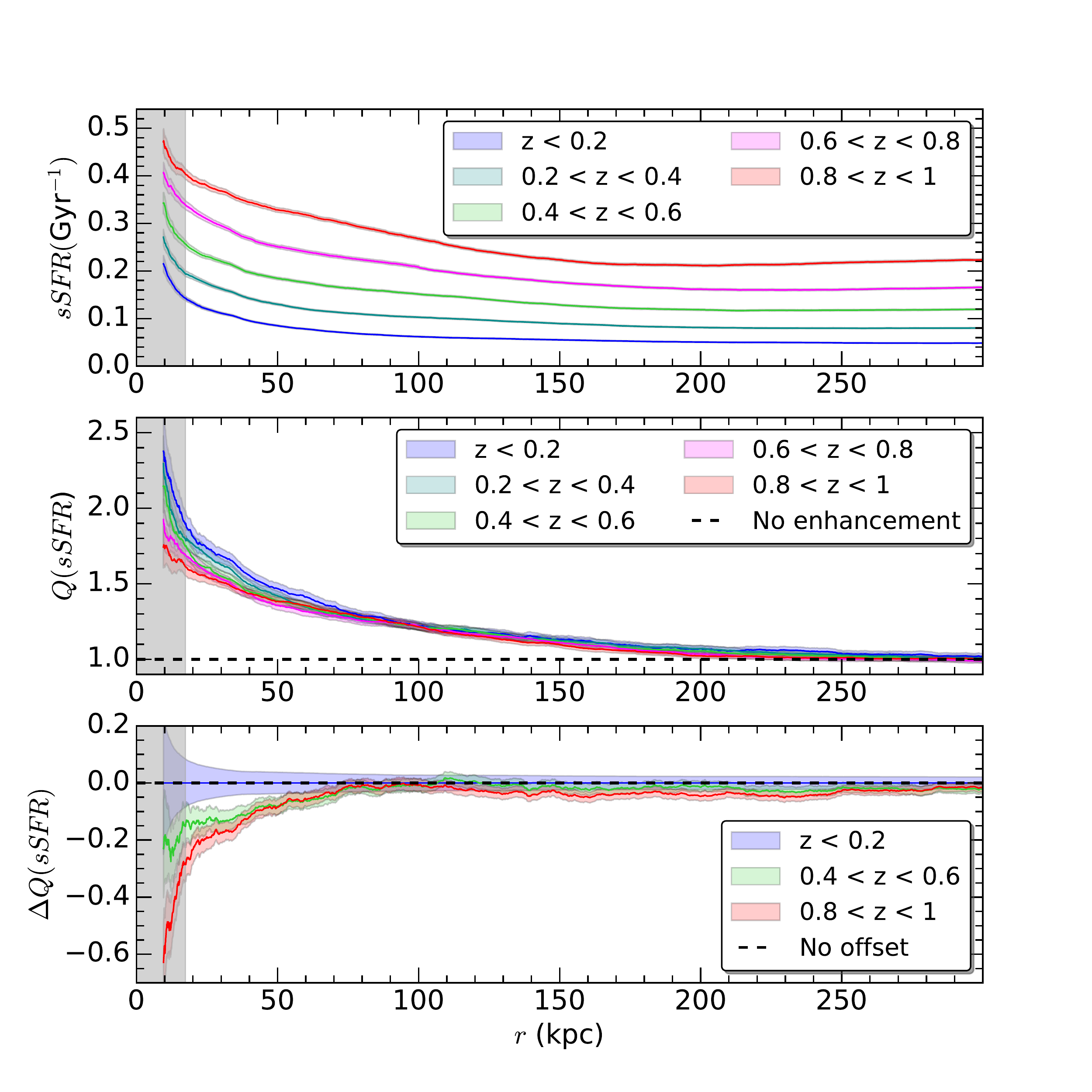}}}}
\caption{In the upper panel, the mean sSFR of TNG300-1 host galaxies is plotted vs. the 3D distance to their closest companion ($r$) for five redshift bins 
spanning $0 \leq z < 1$.  For clarity, the trends of the control galaxies are not shown.
In the middle panel, the mean sSFR enhancement ($Q$) is plotted vs. $r$ for the same redshift bins.
In the lower panel, the offset in $Q$ from the $z < 0.2$ redshift bin is plotted versus $r$ (for clarity, we do not plot 
the $0.2 < z < 0.4$ and $0.6 < z < 0.8$ redshift bins).
In all panels, the shaded coloured regions depict the 2$\sigma$ standard error in the mean, and the light grey shaded region on the left hand side 
depicts the overall $r < r_{\rm crowd}$ region for TNG300-1.
We find similar trends in sSFR enhancements at all redshifts, although the enhancements at small separations 
decrease with increasing redshift.
\label{figt3z}}
\end{figure}

In the upper panel of Fig.~\ref{figt3z}, we plot the mean sSFR of host galaxies 
as a function of $r$ for five redshift bins.  At all separations, 
we find that the mean sSFR increases with redshift.  The mean sSFR increases 
by about a factor of 5 from our lowest redshift bin ($0 \leq z < 0.2$) to our highest redshift bin ($0.8 < z < 1$).
This trend is consistent with redshift-dependent increases in the SFRs of IllustrisTNG galaxies reported 
by the IllustrisTNG collaboration \citep{torrey18,weinberger18,donnari19}, and is qualitatively consistent 
with the observed increases in SFRs with redshift \citep{noeske07,whitaker12,speagle14}.

In the upper panel of Fig.~\ref{figt3z}, we see a similar relationship between the mean sSFR and $r$ within each redshift bin, 
with the mean sSFR clearly rising as $r$ decreases.  In the middle panel of 
Fig.~\ref{figt3z}, we plot the mean sSFR enhancement ($Q$) as a function 
of $r$ for all five redshift bins.  We find that the enhancements in all five redshift bins are 
similar to one another, with statistically significant 
enhancements extending out to at least 200 kpc in every case.
This suggests that interactions cause increased star formation 
throughout the full redshift range of $0 \leq z < 1$, despite strong evolution 
in the underlying sSFRs of galaxies.  Moreover, the fact that similar trends 
are seen in all five redshift bins demonstrates that it is reasonable for us to have combined 
all of the $z < 1$ snapshots together for our analysis.  

Nevertheless, close inspection of the middle panel of Fig.~\ref{figt3z} reveals that 
the sSFR enhancements are somewhat smaller at higher redshifts.
To better visualise this trend, we define a new quantity $\Delta Q$, 
which is the difference between $Q$ in a given redshift bin and $Q$ in the 
lowest redshift bin ($z < 0.2$).  
In the lower panel of Fig.~\ref{figt3z}, we plot $\Delta Q$ versus $r$ for three 
representative redshift bins.  This plot shows that the sSFR enhancements 
are significantly lower at high redshift ($0.8 < z < 1$) 
than at low redshift ($z < 0.2$), especially at $r <  50$ kpc.  
The size of this difference is modest ($\Delta Q \sim 0.3$ at $r = r_{\rm crowd}$) 
but statistically significant.  

This trend is consistent with predictions from simulations that 
SFR enhancements are smaller at higher redshift \citep{perret14,kaviraj15,fensch17,martin17}.
There is also some observational support for this trend \citep[e.g.,][]{kaviraj13,lofthouse17}. 
While a detailed study of the underlying factors driving this trend in IllustrisTNG is beyond 
the scope of this analysis, this topic will be examined in a forthcoming paper.

\section{Comparison with other simulations}

\subsection{Comparison with other IllustrisTNG simulations}\label{secres}

In the previous section, we examined the mean sSFR and its enhancement 
in the TNG100-1 (Fig.~\ref{figt1}) and TNG300-1 (Fig.~\ref{figt3}) simulations.
While similar trends are seen in both, 
we now carry out a direct comparison between these simulations and investigate how the sSFR enhancements 
depend on both the resolution and volume of the simulations.

\begin{figure}
\centerline{\rotatebox{0}{\resizebox{9.0cm}{!}
{\includegraphics{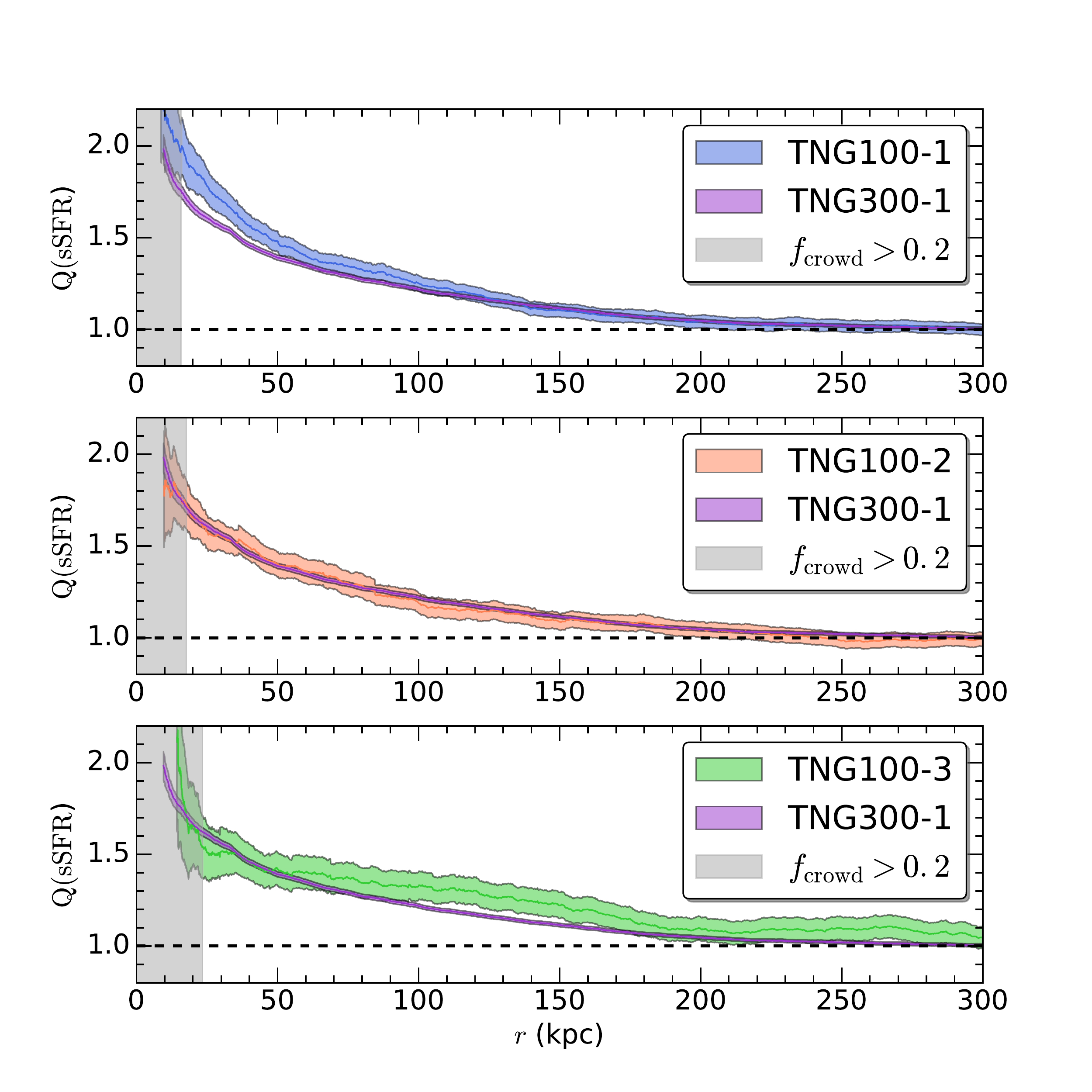}}}}
\caption{sSFR enhancements in TNG300-1 are compared with those in TNG100-1 (upper panel; 8 times higher resolution than TNG300-1), 
TNG100-2 (middle panel; same resolution as TNG300-1) and TNG100-3 (lower panel; 8 times lower resolution than TNG300-1).  
In each panel, the shaded region on the left depicts separations at which $f_{\rm crowd} > 0.2$ for the TNG100 simulation.
While we find similar overall trends at all resolutions, the sSFR enhancements at small separations are larger at higher resolution.
\label{figtng}}
\end{figure}

In Fig.~\ref{figtng}, we plot the sSFR enhancement vs. the distance to the closest companion for the three different 
resolution runs of the IllustrisTNG 110.7 Mpc cube (see Table~\ref{tabsim}).
In each case, we also display the TNG300-1 results for comparison.  In the upper panel of Fig.~\ref{figtng},
we see that while there is generally good agreement between the high resolution TNG100-1 and 
the lower resolution TNG300-1, the sSFR enhancements are consistently higher at $r \lesssim 100$ kpc 
for TNG100-1, with differences significant at the 2$\sigma$ level below 50 kpc. 
It is therefore possible that the sSFR enhancements have not yet converged at small scales 
in TNG100-1. 

In the middle panel of Fig.~\ref{figtng}, we compare two simulations at the same resolution but different volumes 
(TNG100-2 and TNG300-1).
Here, the sSFR enhancements are consistent at all separations, although the $2\sigma$ error in the 
mean is much smaller for the larger volume simulation.  This consistency suggests that the TNG100 box is large 
enough to cover a representative range of environments.  

In the lower panel of Fig.~\ref{figtng}, we compare TNG300-1 with the lowest resolution 100 Mpc simulation (TNG100-3).  
We find that, despite its comparatively poor resolution, TNG100-3 nevertheless captures much of the enhanced star formation 
seen in the larger volume and higher resolution TNG300-1.  The most notable exception lies at small separations, where 
the lower resolution simulation suffers from significant crowding ($f_{\rm crowd} > 0.2$) out to larger separations (23.4 kpc) 
than the other simulations, as expected (see Fig.~\ref{figrsep}).  
This suggests that higher resolution is particularly important in assessing enhanced star formation at smaller separations.

We have found that sSFR enhancements are captured in a relatively well converged fashion across the full TNG100 resolution suite.  
The degree of SFR enhancement is, however, impacted at small separations, with higher resolution simulations capturing 
higher SFR enhancements when galaxies are close.  
These trends are consistent with the findings of \citet{sparre16}, 
who carried out zoom-in simulations of four merging systems from the Illustris simulation over a range of resolutions.  
They report stronger SFR enhancements when running at higher resolution, with differences seen both before and 
after the completion of the merger.   Given that their lower resolution runs are comparable in resolution to TNG100-1, 
this also suggests that the sSFR enhancements we have detected in TNG100-1 are likely to be underestimated.  
This interpretation has additional support from high resolution merger simulations, which indicate that 
resolution plays an important role in capturing star formation in interacting systems \citep{bournaud11,renaud14}.
On the other hand, when comparing high and low resolution runs of the Illustris simulations, 
\citet{sparre15} find similar fractions of star formation occurring in galaxies that reside above the star forming main sequence.  
This suggests that while lower resolution simulations (such as TNG100-3) may perform moderately well for modest increases in star formation, 
they may not be as successful at capturing strong starbursts as their higher resolution counterparts (such as TNG100-1).

\subsection{Comparison with EAGLE and  Illustris-1}

Until now, we have investigated the enhancements in sSFR within the IllustrisTNG suite of simulations, 
exploring how these enhancements depend on resolution, volume and redshift.  
However, these simulations were all produced using the same underlying gravity and hydrodynamics solver, and subgrid models.
To provide a more independent comparison, we now analyse the EAGLE and Illustris-1 simulations to see 
if sSFR enhancements are present and, if so, to compare the size and extent of the sSFR enhancements 
with those in TNG100-1 (which is similar to both EAGLE and Illustris-1 in terms of resolution and volume; see Table~\ref{tabsim}).

\begin{figure}
\centerline{\rotatebox{0}{\resizebox{10.0cm}{!}
{\includegraphics{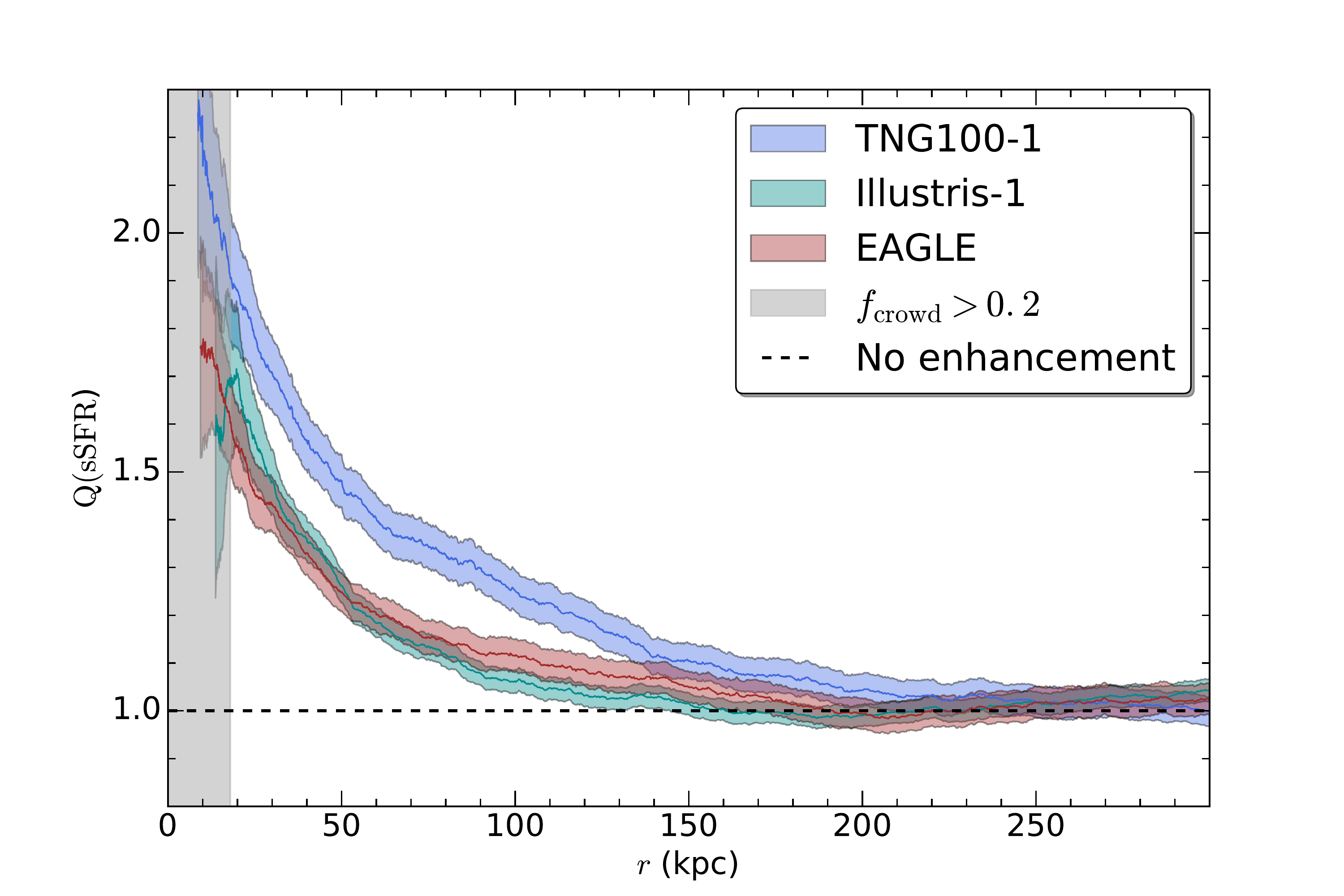}}}}
\caption{The sSFR enhancements in EAGLE, Illustris-1 and TNG100-1 are plotted vs. the distance to the closest companion.
The shaded region on the left depicts the Illustris-1 crowding limit (18.0 kpc), which is slightly 
larger than that of TNG100-1 (15.9 kpc) and EAGLE (16.3 kpc). 
We find similar trends in all three simulations.  However, the sSFR enhancements 
in EAGLE and Illustris-1 are significantly lower than those in TNG100-1.
\label{figeit}}
\end{figure}

We present measurements of sSFR enhancement as a function of $r$ for the EAGLE, Illustris-1 and TNG100-1 simulations in Fig.~\ref{figeit}.
We find that the mean sSFR is also enhanced in EAGLE and Illustris-1.  
The sSFR enhancements in EAGLE rise to $Q\sim1.6 \pm 0.1$ at the crowding limit of 16.3 kpc, with statistically significant (2$\sigma$) 
enhancements extending out to separations of 162 kpc.  
For Illustris-1, the enhancements rise to $Q\sim1.7 \pm 0.1$ at the crowding limit of 18.0 kpc, with significant 
enhancements extending out to 145 kpc.  
Remarkably, the enhancements in EAGLE and Illustris-1 
are consistent with one another (within the $2\sigma$ error bars that are shown) at all separations below 300 kpc.

We also plot the TNG100-1 enhancements (see Sec.~\ref{sect1}) in Fig.~\ref{figeit}.   
The sSFR enhancements in EAGLE and Illustris-1 are significantly lower than those in TNG100-1, by approximately a factor of two.
These differences are statistically significant (at the $2\sigma$ level) below 140 kpc.
The differences between TNG100-1 and Illustris-1 are particularly notable, given that their SFR prescriptions are identical.  
As such, the differences in their sSFR enhancements can likely be attributed to changes in gas properties. 
Given that the sizes and gas fractions of IllustrisTNG and Illustris galaxies are known to be different \citep{pillepich18b}, 
it seems likely that galaxy properties play a central role in determining the level of SFR enhancement.  

While there are some differences in the behaviours of the sSFR enhancements in TNG100-1, EAGLE and Illustris-1, 
what is arguably more striking is the general similarities seen in all three simulation models.  In each case, 
significant enhancements in sSFR are seen, with the strongest enhancements at the smallest separations probed.
In addition, the results are consistent with modest enhancements extending out to $\sim$ 150-200 kpc 
in all three simulations.  This suggests that enhanced star formation due to the presence of a relatively nearby companion 
is a generic outcome for cosmological hydrodynamical simulations, despite the fact that the simulations 
were not tuned in advance to match this relationship.

\section{Connecting cosmological simulations to observations}

We have presented an analysis of how the mean sSFR and its enhancement (with respect to a control sample) depends on the 3D distance 
to the closest companion in various cosmological hydrodynamical simulations.  In order for us to connect these results with observations,  
we now investigate how the results change when 
we quantify proximity in the same way that observers do; namely, using projected 
separation and relative velocity along the line of sight.  We then compare the results 
with a sample of galaxies from the SDSS.

\subsection{Enhanced sSFR vs. projected separation}\label{secrpssfr}

We repeat our analysis from Section~\ref{sec3d}, now using projected quantities instead of 3D separation to quantify proximity.  
Rather than simply converting 3D separations into projected separations, we 
project the sample and then rerun the search for the closest companions, using 
the projected separation ($r_{\rm p}$) and the relative velocity along the line of sight ($\Delta v$) 
to assess proximity.  This more closely mirrors the approach used by observers to identify 
close companions, and accounts for the fact that the closest companion in 3D 
may not be the closest companion in projection \citep{patton00,kitzbichler08,patton08,jian12}.
Moreover, some of the closest companions in this projected sample will be unrelated foreground/background galaxies, 
thereby including a source of contamination familiar to observers.

In principle, we could project the simulations from multiple random orientations in order to 
fully sample the distribution of companion separations.  However, given the 
large volume of the simulations, we elect to simply project the simulations 
in each of the $x-y$, $y-z$ and $x-z$ planes, and use the variation between 
these orthogonal projections to assess the need (if any) for additional projections.

Here we describe our methodology in the context of the $x-y$ projection.  
We take the projected separation $r_{\rm p}$ to be the distance between the host and companion within the $x-y$ plane.  
To match the observational $\Delta v$ criterion along the line of sight, we take the $z$-axis within the cube to be the line of sight, 
and compute $\Delta v$ by summing the contributions from the relative velocity of 
the pair along the $z$-axis and the difference in recession velocities due to the 
Hubble flow.   

Following the methodology that \citet{patton16} used for their observational sample, 
we compute the distance to each host galaxy's closest projected companion ($r_{\rm p}$)
and second closest projected companion ($r_2$) while counting the number of companions within 
a projected separation of 2 Mpc ($N_2$).   We again require potential companions to have stellar 
mass ratios of $\mu_{\rm c} > 0.1$. 
Finally, we require potential companions to have a relative velocity along the line of sight 
(hereafter $\Delta v$) of less than 1000 km s$^{-1}$.  This velocity restriction was used 
by \citet{patton16} to avoid obvious projected companions.

Having identified the closest companion for each host galaxy, we now narrow our analysis to 
host galaxies whose closest companion has $\Delta v<300$ km s$^{-1}$.
This additional restriction on relative velocity increases the likelihood that a given companion 
is physically associated with the host galaxy \citep{patton00}, and corresponds to a maximum 
line of sight separation of 5 Mpc at $z = 0.5$.  Our threshold of 300 km s$^{-1}$ 
was also used by \citet{patton13,patton16}, and is similar to the $\Delta v$ restrictions in many other 
studies of galaxy pairs \citep{lin04,alonso06,lambas12,ellison13,davies15}.

\begin{figure}
\centerline{\rotatebox{0}{\resizebox{9.0cm}{!}
{\includegraphics{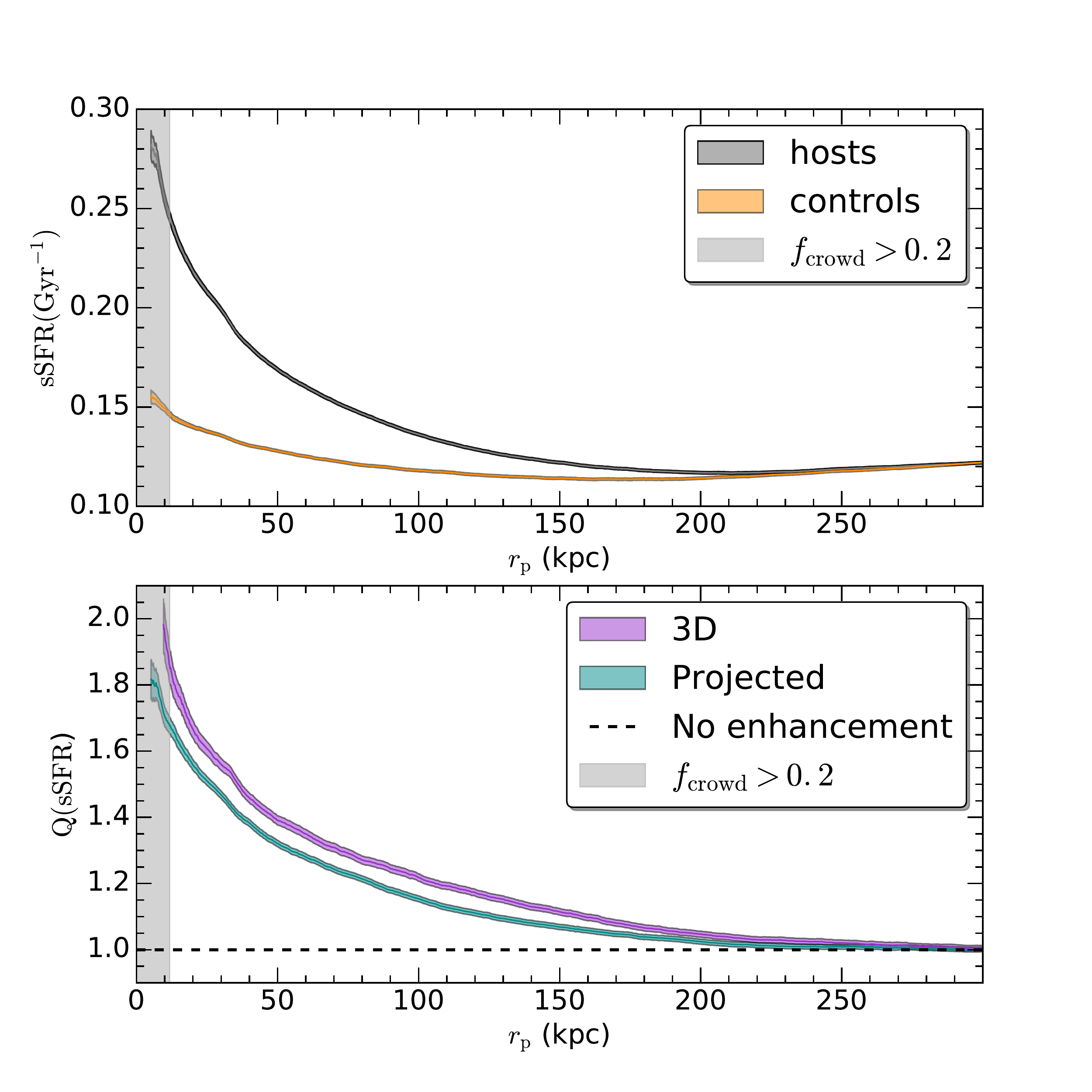}}}}
\caption{In the upper panel, the mean sSFR of TNG300-1 host galaxies and their controls is plotted versus projected separation ($r_{\rm p}$). 
We show only the projection onto the $x-y$ plane in this figure.  
The enhancement in sSFR is plotted versus $r_{\rm p}$ in the lower panel.  
For comparison, we also reproduce measurements of $Q$ versus 3D separation ($r$) from Fig.~\ref{figt3} on the lower panel of this plot.
The shaded regions surrounding each line depict the 2$\sigma$ standard error in the mean.
The region which is susceptible to crowding ($r_{\rm p} < 11.8$ kpc) is depicted in the grey region on the left side of each panel.
We find that projection effects significantly dilute the sSFR enhancements that are seen in 3D.
\label{figrpssfr}}
\end{figure}

In Fig.~\ref{figrpssfr}, we show how the mean sSFR and its enhancement above the control samples 
depends on $r_{\rm p}$  in TNG300-1, 
here only considering the projection onto the $x-y$ plane.
In the upper panel, we plot the mean sSFR of host galaxies and their controls versus the projected distance to 
the host galaxy's closest companion.   The observed trends are qualitatively similar to those seen in 3D (Fig.~\ref{figt3}), 
with a steady increase in the mean sSFR of host galaxies with respect to their controls as the separation decreases.
In the lower panel of Fig.~\ref{figrpssfr}, we plot the mean sSFR enhancement as a function of $r_{\rm p}$.  
This shows that there are statistically significant enhancements (at the $2\sigma$ level) in the mean sSFR at $r_{\rm p} < 260$ kpc, 
with enhancements rising to $Q \sim 1.7$ at the crowding limit of the sample (11.8 kpc).

In the lower panel of Fig.~\ref{figrpssfr}, we also plot the mean sSFR enhancement versus the 3D distance to the closest companion, 
plotting 3D and projected results on the same axis to facilitate a direct comparison between them.  
At a given separation, the enhancements are smaller in projected space than in 3D, implying that 
projection effects are diluting the enhancements that are detected.  At small separations ($r_{\rm p} \lesssim 50$ kpc), 
projection effects dilute the enhancements by about 20 per cent.  At larger separations, the dilution increases 
to the point that enhancements are present in 3D but no longer detectable in projected space.

We can interpret these trends by noting that, at any given sSFR enhancement, $r_{\rm p} < r$.  
This is to be expected, given that a galaxy pair's projected separation 
must be smaller than its 3D separation.  Moreover, in cases where the closest projected companion and the closest 3D companion 
are different galaxies, it is rarely the case that $r_{\rm p} > r$ (for this to happen, the closest 3D companion must be excluded 
from consideration by failing the $\Delta v$ criterion).

We also compare the $x-y$ projection results in Fig.~\ref{figrpssfr}
with projections in the $y-z$ and $x-z$ planes (not shown), and find that they agree within $1\sigma$ at all separations.
This indicates that the TNG300-1 simulation is sufficiently large that the results have been averaged over a representative distribution 
of galaxy pair orientations, thereby mitigating the need to undertake a larger and/or more randomised set of projections.

In summary, we conclude that by measuring sSFR enhancements in projected space (as observers do), 
one is likely to underestimate the underlying level of sSFR enhancements by about 20 per cent at small separations, 
while also underestimating the radial extent of the enhancements.
Nevertheless, the overall similarity of the trends seen in 3D and projected space 
clearly shows that the same underlying correlations produce these relationships.

\subsection{Comparison with Patton et al. (2013)}\label{secp13}

As noted in Section~\ref{secintro}, there is plenty of observational evidence for enhanced star formation in galaxy pairs, 
with the greatest enhancements typically seen in the pairs with the closest projected separations.  
Of particular note is the SDSS study of \citet{patton13}, who use a similar methodology to ours.
They find that the mean SFR is enhanced out to $r_{\rm p} \sim 150$ kpc, 
with the enhancements reaching a factor of $\sim$ 3 in the closest pairs.   
These findings are qualitatively similar to what we have found in TNG300-1 (see Fig.~\ref{figrpssfr}), 
although our enhancements are generally smaller and extend out to larger $r_{\rm p}$.  

However, there are a number of differences between the \citet{patton13} sample and the simulations 
used in this study.  In particular, \citet{patton13} computed mean SFR using a low redshift ($z < 0.2$) 
flux-limited sample that is complete for host galaxies but not companions, whereas we have computed 
mean {\it sSFR} using a $z < 1$ stellar mass-limited sample which is complete for host galaxies and their companions.
We therefore carry out a revised analysis of the \citet{patton13} SDSS sample, addressing most of these 
sample differences.  We acknowledge at the outset that any such comparison 
will be imperfect due to some fundamental differences between the available observational data 
and the quantities reported in the simulations.

\subsection{Comparison with SDSS}\label{secsdss}

We begin with the SDSS sample of star-forming galaxies used by \citet{patton13}, 
and provide here a brief summary of their data and methodology.  
The underlying sample from SDSS \citep{york00} is derived from SDSS Data Release 7 \citep{sdssdr7}, 
using the main galaxy sample described by \citet{strauss02}. 
Total stellar mass estimates are taken from \citet{mendel14} and are based on the photometry of \citet{simard11}.
The sample is restricted to the redshift range $0.02<z<0.2$.

In order to better compare the flux-limited SDSS sample of \citet{patton13} with the stellar mass-limited samples of 
galaxies in the simulations, we apply a minimum stellar mass of $10^{10} \msun$ to host galaxies in SDSS.  
This removes a substantial fraction of the host galaxies used in the analysis of \citet{patton13}, especially at lower redshift.  
In addition, as dwarf galaxies in SDSS exhibit higher SFR enhancements than massive galaxies \citep{stierwalt15}, 
this restriction might be expected to reduce the overall enhancements in our sample. 

Next, in order to maximize the stellar mass completeness of our sample at all redshifts, we impose an additional requirement 
that all galaxies lie above the ``red sequence +3$\sigma$'' redshift-dependent stellar mass limit of \citet{mendel14}.  
This ensures that galaxies of all colours are detectable throughout the sample.  
We then apply ${\rm V}_{\rm max}$ weights to each galaxy, in order to correct for the resulting redshift-dependent mass limits of the sample. 
 To be specific, we apply a statistical weight of $w_{\rm v}$ for each host galaxy, with
\begin{equation}
w_{\rm v} = {d(z_{\rm max})^3 - d(z_{\rm min})^3 \over d(z_{\rm i})^3 - d(z_{\rm min})^3},
\end{equation}
where $d(z)$ is the co-moving line of sight distance at redshift $z$ \citep{hogg99}, 
$z_{\rm min}$ and $z_{\rm max}$ are the survey redshift limits, and $z_{\rm i}$ is 
the redshift at which the sample is complete for galaxies of stellar mass $M_{\rm i}$.
This weight yields the ratio of the co-moving volume at $z_{\rm min} < z < z_{\rm max}$ 
to the available co-moving volume for a host galaxy of stellar mass $M_{\rm i}$. 

Following \citet{scudder12}, we restrict our sample to star-forming galaxies 
using the emission-line criteria of \citet{kauffmann03}.  
We estimate the sSFR of each SDSS galaxy using the fibre SFRs of \citet{brinchmann04} and fibre stellar masses
from \citet{mendel14}.  Given that the typical stellar mass covering fraction of the SDSS fibres for galaxies 
in our SDSS sample is approximately 30 per cent, fibre sSFRs are roughly analogous to the sSFRs computed
within $\rhalf$ of the simulated galaxies used in this study (see Sec.~\ref{secssfr}).
We additionally require galaxies to have sSFR $>$ 0.01 Gyr$^{-1}$, which removes some galaxies which 
are better described as passive than star-forming \citep{brinchmann04,bluck14}.
We note that, while the host galaxies are matched in stellar mass with their control galaxies, averaging of sSFRs rather than SFRs 
will lead to noticeably different results, due to known correlations between SFR and stellar mass \citep{noeske07,peng10}.

We then apply the closest companion and control galaxy methodology of \citet{patton16} 
to this revised SDSS sample.  This includes a weighting scheme that accounts for overall spectroscopic incompleteness 
as well as spectroscopic incompleteness that results from fibre collisions \citep{patton08,patton11,simard11}.
We estimate the enhancement in each host galaxy's sSFR by comparing with at least ten control galaxies, 
assigning higher statistical weights to the controls that provide the best simultaneous matches in redshift, 
stellar mass\footnote{Simultaneous matching on both redshift and stellar mass has the added benefit 
of ensuring that the fibre covering fraction distributions of hosts and controls will be similar to one another \citep{patton11}.}, 
local density and isolation.  We also restrict our sample to host galaxies with closest companions that have $\Delta v < 300$ km s$^{-1}$.
This procedure yields a sample of 141,229 host galaxies from SDSS.

Finally, in order to match the IllustrisTNG simulation to the low redshift star-forming sample from SDSS, 
we apply a maximum redshift of 0.2 to the simulations, and we require a minimum sSFR of 0.01 Gyr$^{-1}$ for host galaxies 
and their controls.

\begin{figure}
\centerline{\rotatebox{0}{\resizebox{10.0cm}{!}
{\includegraphics{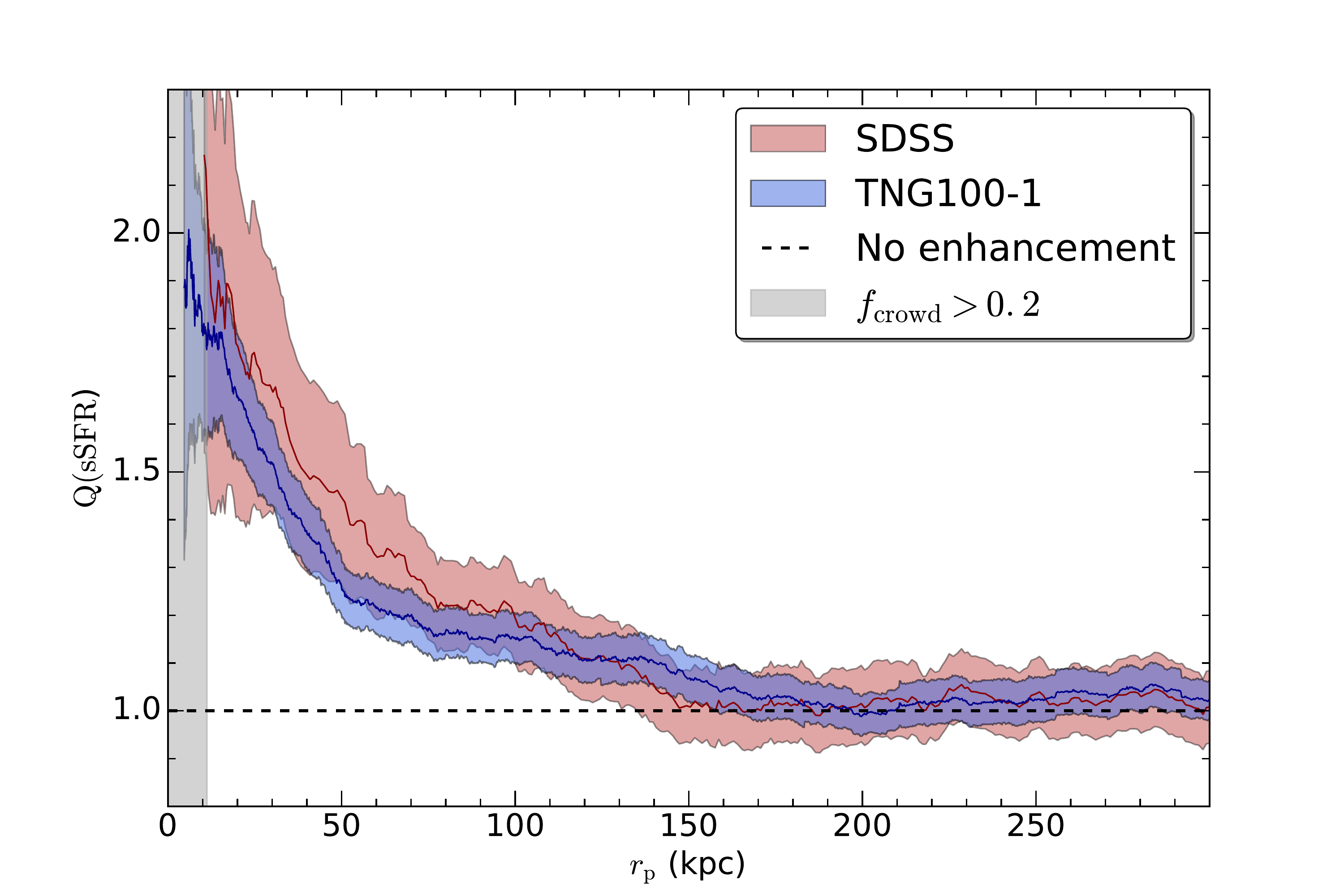}}}}
\caption{sSFR enhancement is plotted vs. projected separation for $z < 0.2$ galaxies from SDSS and TNG100-1.  
Both samples are restricted to star-forming galaxies with sSFR $>$ 0.01 Gyr$^{-1}$.
The coloured shaded regions denote the $2\sigma$ standard error in the mean.
The grey region on the left depicts the separations at which the TNG100-1 sample is susceptible to crowding ($r_{\rm p} < 11.2$ kpc).
\label{figsdss}}
\end{figure}

We plot sSFR enhancement as a function of projected separation for star-forming galaxies 
from SDSS and TNG100-1 in Fig.~\ref{figsdss}.  We find general agreement between the observations and simulations.  In both cases, 
significant sSFR enhancements are seen at $r_{\rm p} \lesssim 150$ kpc,
with the sSFR enhancements increasing to $Q \sim 1.8$ 
at the smallest separations.  The enhancements are somewhat larger in SDSS than in TNG100-1, but 
given their $2\sigma$ error bars, these  differences are not statistically significant.  

These results imply a general consistency between SDSS and TNG100-1. 
However, several caveats are in order.  
First, while our search for companions is complete down to 10 per cent of each host galaxy's mass in IllustrisTNG, 
the same is not true for SDSS; in fact, some SDSS host galaxies lie near the $10^{10} \msun$ limit of the sample.
While it is possible to address this incompleteness by imposing stricter stellar mass limits, 
this removes a substantial fraction of the host galaxies in our SDSS sample, yielding 
results that only poorly constrain the sSFR enhancements and their dependence on $r_{\rm p}$.
Secondly, the typical fibre covering fraction of galaxies in our SDSS sample is about 30 per cent, 
making them more central than the IllustrisTNG sSFRs that are computed within the stellar half mass radius.
Moreover, covering fraction depends on both redshift and galaxy size, leading us to 
average over a wide range of covering fractions.  This means that there are likely to be 
meaningful differences in the spatial coverage of our SDSS and IllustrisTNG samples.  

Nevertheless, the general agreement we have found between SDSS and TNG100-1 suggests that 
the simulations are successfully capturing star formation that is triggered by galaxy-galaxy interactions.  
Moreover, this agreement suggests that the approach we have used for detecting enhanced star formation 
in SDSS is successful at recovering the underlying dependence of sSFR enhancement on the true 
3D separations of galaxies as reported earlier in this paper.

\section{Discussion}

\subsection{Are the sSFR enhancements caused by galaxy-galaxy interactions?}

In Fig.~\ref{figt1}, we discovered statistically significant enhancements in the mean sSFR of host galaxies 
extending out to 3D closest companion separations of 210 kpc in the 
highest resolution IllustrisTNG simulation (TNG100-1).
The much larger sample in the TNG300-1 simulation allowed us to trace significant enhancements 
out even further, reaching 280 kpc (Fig.~\ref{figt3}).  
At these large separations, we would not expect galaxies to be undergoing interactions that are strong enough 
to have any effect on their star formation rates.  However, idealised merger simulations indicate that 
SFR enhancements may be present in galaxies that have recently experienced a close encounter and have 
subsequently moved to a relatively large orbital separation, as long as the SFR enhancement is 
relatively long lived \citep{patton13}.  

In principle, we can test this scenario in IllustrisTNG by tracing out the orbits of interacting galaxies 
to see if  the enhancements we detect at large $r$ are in fact the result of star formation 
triggered by previous close encounters.   However, this would necessitate tracking host galaxies 
and their companions over multiple snapshots.  This is beyond the scope of this paper (which 
uses the properties of host galaxies and their companions at individual snapshots),  
but will be investigated in a forthcoming paper.

Here, we instead investigate whether host galaxies that are experiencing enhanced star formation 
have dark matter haloes that overlap with their closest companions.  
 If so, this would confirm that their dark matter haloes are interacting, and it would increase the likelihood 
that these galaxy pairs have orbits that are decaying due to dynamical friction 
and destined to lead to mergers.
Conversely, if there is little or no overlap between their dark matter haloes, 
it would be harder to attribute the enhanced star formation to 
interactions between these galaxies.

In order to assess the degree to which a host galaxy's dark matter halo overlaps 
with its closest companion, we estimate the virial radius of each, and 
compare the sum of these radii with the separation between the two galaxies.  
We compute the virial radius of the host galaxy ($R_{\rm vir}^{\rm host}$) and its closest companion ($R_{\rm vir}^{\rm comp}$)
using $R_{\rm vir} = 120 (M_{\rm tot}/10^{11}\msun)^{1/3}$ kpc \citep[e.g.,][]{dekel06}, 
where $M_{\rm tot}$ is the total halo mass 
of the galaxy.  
We then compute the relative separation of their dark matter haloes (hereafter $\rsepdm$) as follows: 
\begin{equation}\label{eqnrvir}
\rsepdm = {r \over R_{\rm vir}^{\rm host} + R_{\rm vir}^{\rm comp}}.
\end{equation}
Given this definition, a value of $\rsepdm<1$ means that the host galaxy and its companion have overlapping 
virial radii\footnote{With overlapping haloes, {\textsc SUBFIND} has some trouble accurately deblending the 
dark matter distributions of the two galaxies.  However, we mitigate this effect by 
summing the virial radii of both galaxies in the pair.}.

\begin{figure}
\centerline{\rotatebox{0}{\resizebox{10.0cm}{!}
{\includegraphics{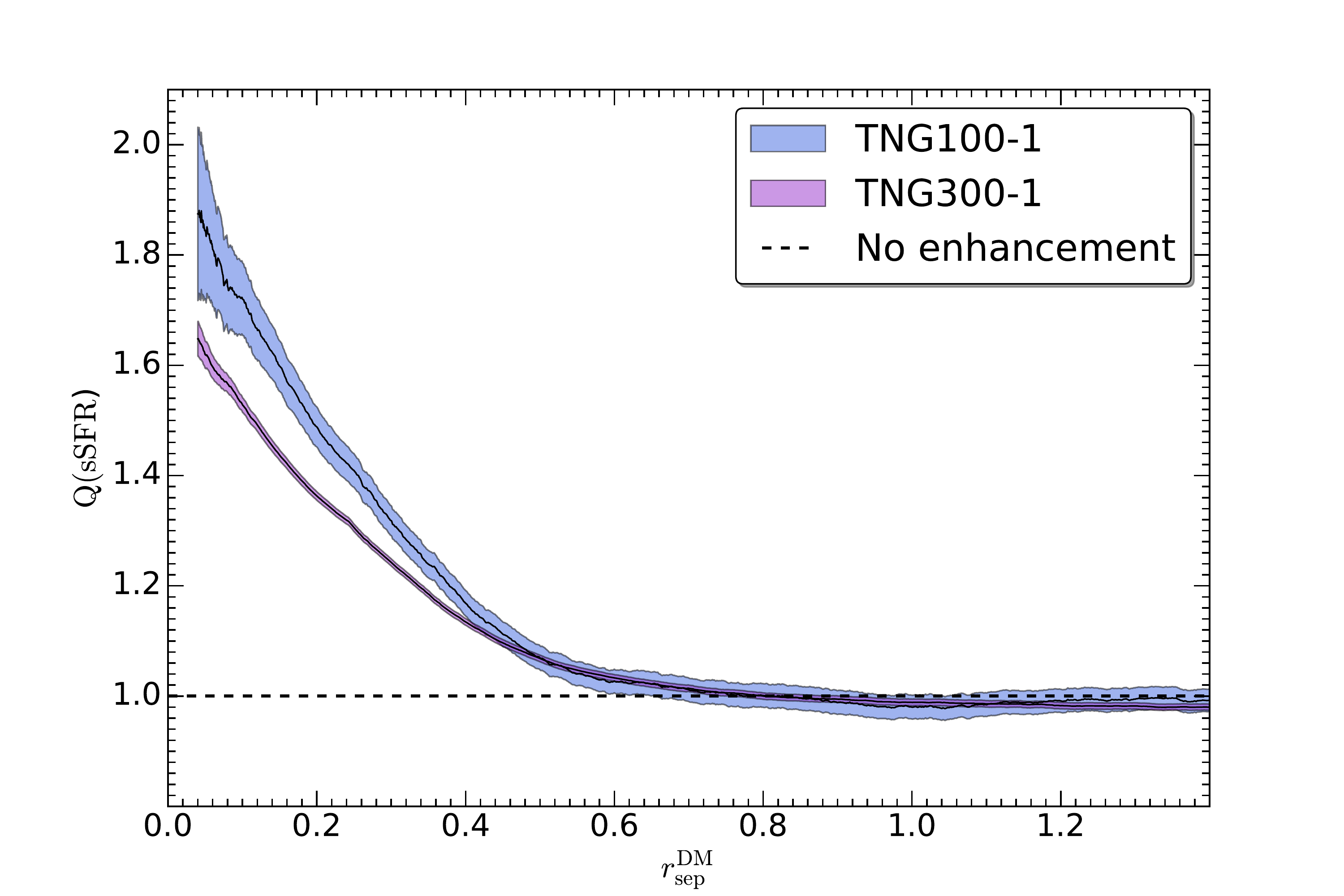}}}}
\caption{Enhanced sSFR is plotted vs. the relative separation of the dark matter haloes ($\rsepdm$) for TNG100-1 and TNG300-1.
The shaded regions depict  the 2$\sigma$ standard error in the mean.
\label{figrvirssfr}}
\end{figure}

We plot the mean sSFR enhancement versus $\rsepdm$ for the highest resolution (TNG100-1) and the largest volume (TNG300-1) IllustrisTNG 
simulations in Fig.~\ref{figrvirssfr}.  We find that the enhancements in mean sSFR are found at $\rsepdm < 1$ for both TNG100-1 and TNG300-1, 
and nearly all of the enhancements occur at $\rsepdm < 0.6$.  This confirms that the sSFR enhancements we have reported, which 
occur for closest companions at $r < 280$ kpc (Fig.~\ref{figt3}) or $r_{\rm p} < 260$ kpc (Fig.~\ref{figrpssfr}), are likely due to 
systems which have substantially overlapping dark matter haloes.   

This finding is consistent with the picture in which the sSFR enhancements are due to galaxy-galaxy interactions.  
The fact that we do not see net enhancements at 
wider separations suggests that post-encounter apocentre separations occur at $\rsepdm < 1$ and/or that 
sSFR enhancements dissipate by the time the galaxies reach a separation of $\rsepdm \sim~ 1$.
However, without tracking sSFR enhancements as a function of orbital properties, 
we cannot distinguish between these scenarios, nor can we say for certain that the enhancements 
are triggered by encounters that are close enough to produce strong gravitational interactions 
between the galaxies.

\subsection{What fraction of cosmic star formation is triggered by interacting galaxies?}

We have detected clear enhancements in the mean sSFR of IllustrisTNG host galaxies out to 
relatively large closest companion separations.  However, while these enhancements 
have high statistical significance, that alone does not mean that the enhancements are 
large enough to have a significant influence on galaxy populations as a whole.  
To address the larger meaning of these results, we need to consider the size of 
these enhancements in conjunction with the fraction of galaxies that are involved.

At small 3D separations, we find that the mean sSFR is enhanced by a factor of 
$2.0 \pm 0.1$ in the highest resolution IllustrisTNG simulation (TNG100-1; see Sec.~\ref{sect1}).
The level of enhancement is slightly lower at small projected separations (see Sec.~\ref{secrpssfr}).
In both cases, these results should be considered lower limits, given that (a) we are unable to 
estimate the enhancements at even smaller separations (due to crowding) 
and (b) the TNG100-1 simulations may not be of high enough resolution to have converged (see Sec.~\ref{secres}). 

These levels of enhancement are broadly consistent with those found in simulations 
and in observations \citep[e.g.][]{cox08,lambas12,patton13,knapen15,martin17,rodriguezmontero19}.
Given that some close pairs contain one or two passive galaxies (potential mixed mergers or dry mergers), 
while others may not be influencing one another (e.g. if they are approaching their first close encounter), 
some of these host galaxies are likely to have sSFR enhancements that are well in excess 
of the mean sSFR enhancement.  

If we consider the full range of closest companion separations at which the mean sSFR of host galaxies 
is enhanced, we can gain a more complete picture of the importance of interactions on galaxy sSFRs.  
To this end, we identify the fraction of the host galaxy population that lies within the regime where the mean sSFR is 
found to be enhanced.   We have previously found that the mean enhancement is 
statistically significant at $r < 280$ kpc in TNG300-1 (Section~\ref{sect3}).  
We find that 37 per cent of host galaxies have $r < 280$ kpc.  This places an upper limit 
on the fraction of the sample affected by these net enhancements.  

More generally, we can examine the impact of enhanced sSFRs on the host sample as a whole by 
computing the mean sSFR of all host galaxies (regardless of the distance to their 
closest companion), and comparing with the mean sSFR of their best controls.  
This is equivalent to replacing every host galaxy with its best control and 
assessing how that would change the mean sSFR of the sample.   
For TNG100-1, we find that the mean sSFR of host galaxies is 0.113 Gyr$^{-1}$, 
versus 0.098 Gyr$^{-1}$ for their best controls.  
This suggests that the presence of closest companions 
(regardless of separation) boosts the mean sSFR of host galaxies by 14.5 per cent.
Approximately one third of this boost (4.7 per cent) comes from host galaxies with close 
companions ($r < 30$ kpc)\footnote{\citet{patton13} reached a similar conclusion when splitting their SDSS pair sample at a projected separation of 30 kpc.}, 
with the remaining two thirds coming almost entirely from host galaxies with closest companions at $30 < r < 200$ kpc.
In other words, while only a small fraction of galaxies have a close companion at a given epoch, 
the combined influence of closest companions at all separations has a notable influence 
on the IllustrisTNG galaxy population as a whole.  
Given the consistency between enhancements in IllustrisTNG and SDSS, 
it is reasonable to infer that the same conclusions may apply to observed galaxy populations at low redshift.

These results are broadly consistent with the findings of \citet{martin17}, who use the Horizon-AGN cosmological hydrodynamical 
simulations \citep{dubois14} 
to estimate the SFR enhancement within a 2 Gyr window surrounding mergers, 
thereby including both interacting pairs and post-mergers.  They report that about 20 per cent of star formation 
at $z \sim 1$ is triggered by mergers, 
with 65 per cent of the enhancement occurring before the merger.  
Conversely, \citet{rodriguezmontero19} find that mergers in the SIMBA simulations \citep{dave19} 
contribute only about 1 per cent to the global SFR budget at $z \sim 2$, 
with this fraction dropping to about 0.5 per cent at $0 < z < 1$.
However, their census refers only to mergers seen shortly after coalescence, 
and does not include star formation that is triggered in the pre-merger phase.

\section{Conclusions}

We have used the IllustrisTNG cosmological hydrodynamical simulations to investigate the relationship 
between enhanced star formation and the presence of close companions.  The primary goal of this 
analysis is to bridge the gap between idealised binary merger simulations (which predict enhanced 
star formation during galaxy-galaxy interactions) and observational studies of galaxy pairs (which 
report enhanced star formation that is associated with the presence of close companions).  
We identify the closest companion for massive galaxies ($M_* > 10^{10} \msun$) 
in various IllustrisTNG simulations, taking care to minimise the
effects of numerical stripping on the estimated stellar masses and sSFRs of galaxies with 
close companions.  
We estimate the enhancement in galaxy sSFRs by comparing with a control sample, using the methodology of \citet{patton16} 
to match on redshift, stellar mass, local density and isolation.   We then analyse how the sSFR enhancement depends on 
the 3D and projected distance to the closest companion, also comparing our results with the Illustris-1 and EAGLE simulations, 
and with a sample of massive galaxies from the SDSS.

Our main conclusions are as follows:
\begin{enumerate}
\item Using the highest resolution IllustrisTNG simulation (TNG100-1; $L_{\rm box} \sim 110$ Mpc), we find that the mean sSFR is 
enhanced by a factor of $2.0 \pm 0.1$ at small separations ($r \sim 16$ kpc; see Fig.~\ref{figt1}), with clear morphological signs of interactions 
seen in synthetic stellar composite images of most of these systems (Fig.~\ref{figt1mosaic}).
\item Using the largest volume IllustrisTNG simulation (TNG300-1; $L_{\rm box} \sim 300$ Mpc), 
we detect statistically significant enhancements in the mean sSFR out to 3D separations of 280 kpc (Fig.~\ref{figt3}).
\item Clear enhancements in the mean sSFR are seen throughout the redshift range $0 \le z < 1$ for TNG300-1, although there is 
a gradual decrease in small scale ($r < 50$ kpc) sSFR enhancements as redshift increases (Fig.~\ref{figt3z}).
\item Similar sSFR enhancements are detected across the full range of TNG100 resolutions, although 
the highest resolution simulation (TNG100-1) exhibits larger enhancements below 50 kpc (Fig.~\ref{figtng}).
\item We also detect significant sSFR enhancements in the Illustris-1 and EAGLE simulations, 
although the enhancements are smaller (by about a factor of two) and extend out to smaller separations than in TNG100-1 (Fig.~\ref{figeit}).
\item After redefining our closest companion sample using projected separation, we find that projection effects dilute the 
TNG300-1 sSFR enhancements by about 20 per cent for close companions, and they narrow the range of separations at which 
enhancements are detected (Fig.~\ref{figrpssfr}).
\item Using low redshift star forming samples of galaxies from TNG100-1 and SDSS, we find general agreement 
in sSFR enhancements between simulations and observations (Fig.~\ref{figsdss}).
\item The sSFR enhancements in IllustrisTNG occur in systems where the host galaxy and its closest companion have overlapping 
virial radii, confirming that these enhancements are associated with interacting galaxies (Fig.~\ref{figrvirssfr}).
\item By summing the sSFR enhancements at all separations, we estimate that closest companions boost the mean sSFR of massive galaxies 
in TNG100-1 by 14.5 per cent.
\end{enumerate}

We have shown that cosmological hydrodynamical simulations are able to capture 
enhanced star formation associated with galaxy-galaxy interactions, exhibiting  
general consistency with high resolution merger simulations and observations of galaxy pairs.  
\new{In the second paper in this series, \citet{hani20} report that star formation is also enhanced 
in IllustrisTNG post-merger galaxies.}
These findings provide motivation for using IllustrisTNG to assess the influence of interactions 
on additional galaxy properties, such as metallicities, morphologies, and gas content.
Moreover, the ability to track the orbital history and future of interacting galaxies in these simulations 
will enable us to develop a deeper understanding of the 
physical processes that give rise to these changes in galaxy properties.

\section*{Acknowledgements}

We thank the anonymous referee for a thoughtful and constructive referee's report.
We thank all members of the IllustrisTNG, Illustris, EAGLE and SDSS collaborations for making their data available.
DRP and SLE gratefully acknowledge NSERC for Discovery Grants which helped to fund this research.  
CM, KW and WB were supported by NSERC Undergraduate Student Research Awards.
Support for JM was provided by the NSF (AST Award Number 1516374), and by the Harvard Institute for Theory and Computation, 
through their Visiting Scholars Program.
MHH acknowledges the receipt of a Vanier Canada Graduate Scholarship.
We thank Connor Bottrell for helpful discussions which improved this manuscript.  

The EAGLE simulations were performed using the DiRAC-2 facility at Durham, 
managed by the ICC, and the PRACE facility Curie based in France at TGCC, CEA, Bruy\`eres-le-Ch\^atel.

Funding for the SDSS and SDSS-II has been provided by the Alfred P. Sloan Foundation, the Participating Institutions, the National Science Foundation, the U.S. Department of Energy, the National Aeronautics and Space Administration, the Japanese Monbukagakusho, the Max Planck Society, and the Higher Education Funding Council for England. The SDSS Web Site is http://www.sdss.org/.

The SDSS is managed by the Astrophysical Research Consortium for the Participating Institutions. The Participating Institutions are the American Museum of Natural History, Astrophysical Institute Potsdam, University of Basel, University of Cambridge, Case Western Reserve University, University of Chicago, Drexel University, Fermilab, the Institute for Advanced Study, the Japan Participation Group, Johns Hopkins University, the Joint Institute for Nuclear Astrophysics, the Kavli Institute for Particle Astrophysics and Cosmology, the Korean Scientist Group, the Chinese Academy of Sciences (LAMOST), Los Alamos National Laboratory, the Max-Planck-Institute for Astronomy (MPIA), the Max-Planck-Institute for Astrophysics (MPA), New Mexico State University, Ohio State University, University of Pittsburgh, University of Portsmouth, Princeton University, the United States Naval Observatory, and the University of Washington.





\bsp	
\label{lastpage}
\end{document}